\documentclass[12pt]{article}
\usepackage{graphicx}
\usepackage{newtxtext,nicefrac}
\usepackage{newtxmath,xcolor,tikz}
\usepackage{hyperref}
\usepackage[skip=0pt]{caption}
\usepackage[skip=0pt]{subcaption}
\hypersetup{
    colorlinks = true,
    urlcolor   = blue,
    citecolor  = black,
}

\newcommand{\RomanNumeralCaps}[1]
\linenumbers
\allowdisplaybreaks
\newcommand{\bseq}{\begin{subequations}}
\newcommand{\eseq}{\end{subequations}}
\newcommand{\beq}{\begin{equation}}
\newcommand{\eeq}{\end{equation}}
\newcommand{\bef}{\begin{figure}}
\newcommand{\eef}{\end{figure}}
\newcommand{\blackline}{\raisebox{2pt}{\tikz{\draw[-,black,solid,line width = 1.5pt](0,0) -- (5mm,0);}}}
\newcommand{\greenline}{\raisebox{2pt}{\tikz{\draw[-,green,solid,line width = 1.5pt](0,0) -- (5mm,0);}}}
\newcommand{\blueline}{\raisebox{2pt}{\tikz{\draw[-,blue,solid,line width = 1.5pt](0,0) -- (5mm,0);}}}
\newcommand{\cyanline}{\raisebox{2pt}{\tikz{\draw[-,cyan,solid,line width = 1.5pt](0,0) -- (5mm,0);}}}

\begin{document}

\title{Spatiotemporal linear instability of viscoelastic slender jets}
\author{ {\bf T. Chauhan$\dagger$}, {\bf D. Bansal$\dagger$} and {\bf S. Sircar$\dagger, \ddagger$}
\\{$\dagger$ \small Department of Mathematics, IIIT Delhi, India 110020} \\{\small $\ddagger$ Corresponding Author (email: sarthok@iiitd.ac.in)}}
\maketitle
\date{ }

\begin{abstract}
We revisit the problem of the two-dimensional spatiotemporal linear stability of viscoelastic slender jets obeying linear Phan-Thien-Tanner (PTT) stress constitutive equation, and investigate the role of finite stresses on the elasto-capilliary stability of the Beads-on-a-String (BOAS) structure (structures which include the formation of very thin filament between drops) and identify the regions of topological transition of the advancing jet interface, in the limit of low to moderate Ohnesorge number ($Oh$) and high values of Weissenberg number ($We$). The Briggs idea of analytic continuation \{previously elucidated in the nonaffine response regime [Bansal, Ghosh and Sircar, ``Spatiotemporal linear stability of viscoelastic free shear flows: Nonaffine response regime'', Phys. Fluids {\bf 33}, 054106 (2021)]\} is deployed to classify regions of temporal stability and absolute and convective instabilities, as well as evanescent modes, and the results are compared with previously conducted experiments for viscoelastic filaments. The impact of the finite stresses are evident in the form of strain-hardening ensuing in lower absolute growth rates, relatively rapid drainage and finite-time pinch-off. The phase diagrams reveal the influence of (a) capillary force stabilization at infinitesimally small values of $Oh$, and (b) inertial stabilization at significantly larger values of $Oh$.
\end{abstract}
\maketitle

\noindent {\bf Keywords:} Phan-Thien and Tanner model, Spatiotemporal stability, Beads-On-A-String, topological transition

\section{Introduction}\label{sec:intro}
The instability and the subsequent disintegration of a slender column of fluid, continues to be an active area of research in a wide range of scientific disciplines and in emerging applications such as ink jet printing, nanofibers, needle-free injections~\cite{Moradiafrapoli2017}, coating and diesel engine technology~\cite{Turner2012}. Liquid jets (defined as a stream which is propelled into a medium through an outlet such as a nozzle) are unstable in nature and breakup into droplets~\cite{Eggers1993, Eggers1994JFM} and the primary mechanism of this breakup is the Rayleigh-Plateau instability which is identified by the growth of disturbances that are either absolutely or convectively unstable~\cite{Brenner1994, Brenner1996, Brenner1997}. While convective instability grows in amplitude as it is swept along by the flow, absolute instability occurs at fixed spatial locations, leading to pinch-off~\cite{Gallaire2017} and recoil~\cite{Chang1999}. Furthermore, droplet formation can occur either directly at the jet exit or further downstream, at the end of the liquid jet. These two types of instabilities are referred to as dripping and jetting, respectively~\cite{Sunol2015}. 

For the case of Newtonian jets, while the primary stages of destabilization is governed via surface tension, the final nonlinear regime of droplet pinch-off represents a finite time singularity and is described by the use of the so-called self-similarity solution~\cite{Eggers1993}. In contrast, the hydrodynamics of non-Newtonian jets differ from their Newtonian counterparts; since elasticity drastically changes the character of the surface instabilities~\cite{Entov1997}. When small concentrations of flexible polymer is added to a solvent, pinch-off is strongly delayed, the liquid jet becomes sufficiently thin (in the so called `stretching stage') and, in the time-asymptotic (or within the `elastic drainage'~\cite{Chang1999}) regime, a viscoelastic Beads-on-a-String BOAS configuration is formed between the nozzle and the droplet~\cite{Bhat2010}. In the drainage regime, the filament radius reaches a constant value, the axial elastic stress is large and the strain rate drops from a unit order to a negligibly small value, such that there is no flow out of the filament due to stretching~\cite{Ardekani2010}. The BOAS configuration is established due to a significant change in the magnitude of the elastic stresses and a distinctively different balance (compared with the one found in the stretching stage) between the capillary, viscous, elastic and inertial forces~\cite{Clasen2006}.

A plethora of recent theoretical~\cite{Turkoz2018,Alhushaybari2019} and computational~\cite{Zinelis2019, Panahi2020} analyses have focused on the breakup dynamics of slender jets via the Oldroyd-B or the Upper Convected Maxwell (UCM) models, leading to simplifications where the polymer chains are assumed to be infinitely stretchable (equivalently, fluids exhibiting infinite stress). The choice of these models imply that the jet does not break up in finite time. Other groups have mitigated this unphysical behavior by choosing the Finitely Extensible Nonlinear Elastic constitutive equation (FENE-P)~\cite{Ferreira2016,Shamardi2018,Guimaraes2020}, although with restricted models, especially avoiding the impact of the radial stresses on the evolution of the jet interface.

We have limited our focus on the linear spatiotemporal analyses of viscoelastic slender jets obeying the linear Phan Thien Tanner (PTT) model, to identify the material parameter regions leading to pinch-off. We characterize the fluid flow as being absolutely or convectively unstable (while simultaneously detecting the Evanescent (false) modes~\cite{Patne2017}), since convective instability allows the disturbance to pass through the jet filament without triggering pinch-off. The method of spatiotemporal analysis by progressive moving of the isocontours in the complex frequency and wavenumber plane, as proposed by Kupfer~\cite{Kupfer1987}, is utilized. The present work significantly differs from existing studies~\cite{Sunol2015} in the sense that we analyze the impact of both the finite axial as well as the finite radial stresses on the linear spatiotemporal stability of viscoelastic slender jets and aim to address the following intriguing questions: What is the critical flow/polymer relaxation condition for the onset of instability? and more crucially, what is the linear spatiotemporal, time asymptotic response of the flow at the critical value of the material parameters, leading to the topological transition of the advancing interface? The next section (\S \ref{subsec:GE}) outlines the slender body model of the viscoelastic jet flow coupled with the base constitutive relation for the extra elastic stress tensor: linear PTT model. \S \ref{subsec:MF} delineates a quasi-stationary, time-asymptotic solution of the model, whose linear stability is described in \S \ref{subsec:LA}. \S \ref{subsec:NM} outlines the choice of the numerical method in tracking the absolute growth rate curves and the material parameter boundaries of the spatiotemporally unstable regions. \S \ref{sec:Results} showcases the simulation results including the spatiotemporal stability analyses (\ref{subsec:STI}), followed with a brief discussion on the implication of these results as well as the focus of our future direction (\S \ref{sec:Conclusions}). The appendix (\S \ref{sec:appendix}) lists all the coefficients of the Dispersion Relation.

\section{Problem formulation and numerical method}\label{sec:math}
In this paper we seek the spatiotemporal stability of the (temporally) asymptotic solution that encompasses the inertio-elasto-capilliary balance, observed in the experiments of liquid bridges, pinch drops and thinning jets~\cite{Bhat2010}. We follow the spirit of some of the earlier work by employing two simplifying assumptions~\cite{Keshavarz2016}. First, we are treating the flow inside the fluid thread as effectively one-dimensional~\cite{Eggers1994PHF}. This simplification is consistent as long as the shape of the liquid column remains slender, i.~e., the characteristic radial variations are small compared to the variation in the axial direction. This assumption is problematic near the ends of the fluid drops in the BOAS structure~\cite{Eggers1994JFM}.

Next, we will restrict our studies in the limit $We \gg 1$, implying that the non-Newtonian polymer contribution is significant at all times. Thatis, the polymers become sufficiently stretched to counter surface tension forces and the simplified, local system of equations converges to a `quasi-stationary' solution, maintained by the stress in the polymers with no possibility of relaxation~\cite{Chang1999}. In particular, the role of gravity is negligible in this limit.

Although it is an experimentally established fact~\cite{Bhat2010} that low (but finite) inertia is necessary to induce the formation of alternating beads-filament structure, certain simulations have reported the bead-formation in the absence of inertia, when the fluid was modeled using the PTT network model, incorporating non-affine motion~\cite{Matallah2006} and in strong transient flows with discrete (or step) shear stress~\cite{Bhat2009}. Hence, these anomalous cases have been avoided in the present study.

\subsection{Governing equations}\label{subsec:GE}
We consider the linear stability of an axisymmetric, quasi-steady, fully developed flow of a vertically falling, viscoelastic, slender jet which is thinning away from the inlet (see figure~\ref{fig1}). A cylindrical coordinate system is used with $r$ and $z$ denoting the radial and the axial direction, respectively. 
\bef
\centering
\includegraphics[width=0.4\linewidth, height=0.4\linewidth]{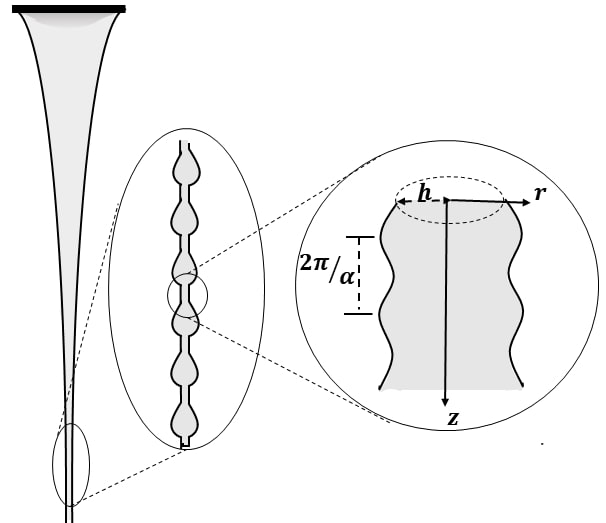}
\caption{Schematic diagram of an axisymmetric surface instability (of wavelength $\nicefrac{2\pi}{\alpha}$) of a viscoelastic liquid jet of radius $h$, highlighting two distinct regimes of flow-transition: (a) the stretching regime of the thinning jet and (b) the elastic drainage regime leading to the Beads-On-A-String (BOAS) configuration.}
\label{fig1}
\eef
The following scales are used for non-dimensionalizing the governing equations: the initial undisturbed radius of the jet at the start of the formation of the beads on a string structure, $h(0)$ for lengths, $\dfrac{\rho {h(0)}^2}{\eta_0}$ for time and $\dfrac{\eta^2_0}{{h(0)}^2 \rho}$ for pressure and stresses, with density and the viscosity of the fluid being $\rho$ and $\eta_0=\eta_s + \eta_p$ ($\eta_s, \eta_p$ are the solvent and the polymeric contribution to the shear viscosity), respectively. The governing (non-dimensional) continuity and momentum equations are given by,
\beq
\nabla \cdot {\bf v} = 0, \qquad Re\left[ \frac{\partial {\bf v}}{\partial t} + ({\bf v} \cdot \nabla) {\bf v} \right] = -\nabla p + \nabla \cdot \tau,
\label{eqn:momentum}
\eeq
where ${\bf v} = (v_r, v_z)$ are the radial and the axial components of the velocity and $p$ is the pressure. The (non-dimensional) extra stress tensor, $\tau$, is given as follows,
\beq
\tau = \nu {\bf D} + (1-\nu) {\bf A},
\label{eqn:tau}
\eeq
where the parameter, $\nu = \dfrac{\eta_s}{\eta_s + \eta_p}$, represents the viscous contribution to the total viscosity, $D = \nabla {\bf v} + \nabla {\bf v}^T$ is the shear rate tensor and {\bf A} is the elastic contribution to the stress tensor, satisfying the linear PTT equation~\cite{PTT1977,Bansal2021},
\beq
\frac{\partial {\bf A}}{\partial t} + {\bf v}_2 \cdot \nabla {\bf A} -  \nabla{\bf v}_2^T\cdot {\bf A} - {\bf A} \cdot \nabla {\bf v}_2 = \frac{{\bf D} - (1 + \epsilon\,\,We\,\,tr({\bf A})){\bf A}}{We}.
\label{eqn:EStress}
\eeq
In equation~\eqref{eqn:EStress}, $\epsilon \in [0, 1]$ is a dimensionless parameter describing the maximum elongation of the polymer chains. The appropriate boundary conditions are the balance of the normal stress,
\beq
{\bf n} \cdot \tau \cdot {\bf n} = -\dfrac{1}{Oh^2} \kappa,
\label{eqn:NStress}
\eeq
and the tangential stress,
\beq
{\bf n} \cdot \tau \cdot {\bf t} = 0,
\label{eqn:TStress}
\eeq
at the jet interface defined by $r = h(t, z)$. Here ${\bf n}$ is the outward normal and $\kappa$ is the curvature of the interface. Finally, there is also the kinematic condition for mass conservation,
\beq
\dfrac{\partial h}{\partial t} + v_z \dfrac{\partial h}{\partial z} = v_r|_{r = h}.
\label{eqn:Kinematic}
\eeq
The dimensionless parameters involved in the model, equations~(\ref{eqn:momentum}-\ref{eqn:Kinematic}) are the Ohnesorge number, $Oh = \sqrt{\dfrac{\eta^2_0}{\rho \gamma h(0)}}$ (or the ratio of the viscous to the inertio-capillary forces), the Weissenberg number, $We = \dfrac{\lambda \eta_0}{\rho {h(0)}^2}$ (or the ratio of the elastic to the viscous forces), and the Reynolds number, $Re = \dfrac{\eta^2_0}{Oh^2 \gamma \rho h(0)}$ (or the ratio of the inertial to the viscous forces). $\gamma$ and $\lambda$ are the surface tension of the liquid-air interface and the polymer relaxation time, respectively. Following Eggers~\cite{Eggers1993}, we assume a slender jet approximation, expand the variables in the model~(\ref{eqn:momentum}-\ref{eqn:Kinematic}) into a power series in $r$, and arrive at the following dimensionless equations in the long wave limit (where it is assumed that $h(z, t)$ varies slowly with respect to $z$, the pressure, the stress components, $A_{rr}$ and $A_{zz}$ and axial velocity, $v_z = u$, are almost uniform with respect to $r$ while the radial velocity $v_r$ and the off-diagonal stress components, $A_{rz}$ and $A_{zr}$ are nearly zero), 
%
%
%
\bseq \label{eqn:SJmodel}
\begin{align}
& Re \left[ \frac{\partial u_0}{\partial t} + u_0 \frac{\partial u_0}{\partial z} \right] = \frac{1}{Oh^2}\frac{\partial \kappa}{\partial z} + \frac{1-\nu}{h_0^2}\frac{\partial }{\partial z}\left[h_0^2 (A_{zz_0}-A_{rr_0})\right] + \frac{3\nu}{h_0^2}\frac{\partial }{\partial z} \left[h_0^2 \frac{\partial u_0}{\partial z}\right], \label{eqn:SJmomentum} \\
& \frac{\partial h_0^2}{\partial t} + \frac{\partial}{\partial z} (h_0^2 u_0) = 0, \label{eqn:SJkinematic} \\
& \frac{\partial A_{rr_0}}{\partial t} + u_0 \frac{\partial A_{rr_0}}{\partial z} + A_{rr_0}\frac{\partial u_0}{\partial z} = -\frac{1}{We} \left\{ \frac{\partial u_0}{\partial z} + (1 + \epsilon We(A_{rr_0}+A_{zz_0})) A_{rr_0} \right \}, \label{eqn:SJArr} \\
& \frac{\partial A_{zz_0}}{\partial t} + u_0 \frac{\partial A_{zz_0}}{\partial z} -2 A_{rr_0}\frac{\partial u_0}{\partial z} = -\frac{1}{We} \left \{ -2 \frac{\partial u_0}{\partial z} + (1 + \epsilon We(A_{rr_0}+A_{zz_0})) A_{zz_0} \right \}, \label{eqn:SJAzz}
\end{align}
\eseq
where the subscript `0' represents the leading order approximation of the variables with respect to $r$. The jet curvature satisfies the relation,
\beq
\kappa = \frac{h_{zz}}{(1 + h^2_z)^{3/2}} - \frac{1}{h(1 + h^2_z)^{1/2}}.
\label{eqn:curvature}
\eeq
%

\subsection{Mean flow}\label{subsec:MF}
The breakup mechanics of slender viscoelastic filaments is known follow two complex stages of flow-transition before an eventual pinch-off~\cite{Clasen2006}: the stretching stage (at early times) followed by a (late stage) elastic drainage regime. In this study, we will focus on the effect of the extensional flow on the jet filament.

In the early elastic time regime, $t \ll 1 (\ll We)$, elastic stresses are small and elasticity has negligible effect on the initial evolution of the jet. The axisymmetric extensional flow triggers a stretching evolution that enlarges the length of the filament, with locally constant radius, a linear axial velocity and a constant positive axial stress, $A_{zz}$ until a straight filament is formed~\cite{Chang1999}. Anticipating the length of stretching stage to be governed predominantly by the slow extension flow, the fluid inertia is expected to be negligible and the curvature $\kappa$ (equation~\eqref{eqn:curvature}) to be well approximated by the azimuthal curvature only, $\kappa = -1/h$. The dominant terms in the model~\eqref{eqn:SJmodel} during the filament stretching stage are
\bseq \label{eqn:Stage1model}
\begin{align}
& \frac{1}{h_0^2} \frac{\partial }{\partial z} \left[ \frac{h_0}{Oh^2} + (1-\nu) h_0^2 (A_{zz_0}-A_{rr_0}) + 3\nu h_0^2 \frac{\partial u_0}{\partial z} \right] = 0, \label{eqn:Stage1momentum} \\
& \frac{d}{dt} h_0^2 = -\frac{\partial u_0}{\partial z} h_0^2,  \label{eqn:Stage1kinematic} \\
& \frac{d}{dt} A_{rr_0} = - \frac{\partial u_0}{\partial z} \left( A_{rr_0} + \frac{1}{We} \right) - \frac{A_{rr_0}}{We} - \epsilon (A_{rr_0} + A_{zz_0}) A_{rr_0}, \label{eqn:Stage1Arr} \\
& \frac{d}{dt} A_{zz_0} = 2 \frac{\partial u_0}{\partial z} \left( A_{zz_0} + \frac{1}{We} \right) - \frac{A_{zz_0}}{We} - \epsilon (A_{rr_0} + A_{zz_0}) A_{zz_0}, \label{eqn:Stage1Azz}
\end{align}
\eseq
where the maximum radius, $h$, and the elastic stresses, $A_{rr}$ and $A_{zz}$, in the kinematic and the stress equations~(\ref{eqn:Stage1kinematic}-\ref{eqn:Stage1Azz}), are assumed to evolve along the characteristic lines: $\frac{d z}{d t} = u$ (and hence $\frac{d}{dt}(\cdot) = \left[ \frac{\partial }{\partial t} + u \frac{\partial }{\partial z} \right] (\cdot) $ in equations~(\ref{eqn:Stage1kinematic}-\ref{eqn:Stage1Azz})). The hyperbolic nature of the equations~(\ref{eqn:Stage1kinematic}-\ref{eqn:Stage1Azz}) originate from the fact that the liquid mass is convected downstream by the axial velocity. 

Next, we integrate equations~(\ref{eqn:Stage1Arr}, \ref{eqn:Stage1Azz}) to obtain the elastic stresses on the filament. The linear stability analysis by Chang~\cite{Chang1999} using the Maxwell's model in the stretching regime, suggested that the elastic stresses on the filament are a factor of $We^{-1}$ smaller than $h$ and $(\frac{\partial u}{\partial z})$. We couple this information along with the fact that the elastic stresses are small at this stage, ensuing in the following set of equations governing the approximate evolution of stresses on the filament,
\bseq \label{eqn:Stage1RStress}
\begin{align}
& \frac{d}{dt} A_{rr_0} = - \frac{\partial u_0}{\partial z} \left( A_{rr_0} + \frac{1}{We} \right), \label{eqn:RSArr} \\
& \frac{d}{dt} A_{zz_0} = 2 \frac{\partial u_0}{\partial z} \left( A_{zz_0} + \frac{1}{We} \right), \label{eqn:RSAzz}
\end{align}
\eseq
which is solved together with the kinematic condition~\eqref{eqn:Stage1kinematic} and the initial conditions (at $t=0$: $h_0=1, A_{rr_0} = A_{zz_0} = 0$), to arrive at the solution,
\bseq \label{eqn:Stage1Asolution}
\begin{align}
& A_{rr_0} = \frac{h_0^2 - 1}{We}, \label{eqn:ArrSolution} \\
& A_{zz_0} = \frac{h_0^{-4} - 1}{We}, \label{eqn:AzzSolution}
\end{align}
\eseq

Similarly, an elementary integration of equation~\eqref{eqn:Stage1momentum} with respect to $z$, yields the following quasi-steady force balance,
\beq
\frac{h_0}{Oh^2} + (1-\nu) h_0^2 (A_{zz_0}-A_{rr_0}) + 3\nu h_0^2 \frac{\partial u_0}{\partial z} = \zeta(t).
\label{eqn:Stage1FB}
\eeq
which reveals how the local flow gradient is ascertained by the local capillary pressure and the elastic stress difference. Numerical simulations reveal that the force, $\zeta(t)$, does not vary significantly during this stretching interval~\cite{Fontelos2003} and (in essence) can be found via the initial conditions, such that $\zeta(t) \approx \frac{1}{Oh^2}$. Substituting this value in equation~\ref{eqn:Stage1FB}, leads to the following relation
\beq
\frac{\partial u_0}{\partial z} = \frac{1-h_0}{3\nu h_0^2 (Oh^2)} + \frac{1-\nu}{3\nu}\left[ A_{rr_0}-A_{zz_0} \right].
\label{eqn:Stage1uz}
\eeq
Equation~\ref{eqn:Stage1uz}, combined with the kinematic condition~\eqref{eqn:Stage1kinematic} and the solution set~\eqref{eqn:Stage1Asolution}, gives rise to the following differential equation governing the thinning rate on the filament,
\beq
h_0^2 \frac{d h_0^2}{d t} = -\frac{(h_0^2 - h_0^3)}{3 \nu (Oh^2)} + \frac{(1-\nu)}{3\nu}\frac{(1-h_0^6)}{We},
\label{eqn:Stage1h}
\eeq
where the right side represents the flow gradient as driven by the azimuthal curvature difference and retarded by the elastic stress difference. This stretching of the filament ceases when the capillary pressure increases sufficiently as $h$ decreases to balance the stress difference in equation~\eqref{eqn:Stage1h}, leading to a steady-state value of the thinning jet radius (found after ignoring the coefficients of $h_0^3$ and other higher powers of $h_0$ in equation~\eqref{eqn:Stage1h}),
\beq
h^* = Oh \sqrt{\frac{1-\nu}{We}}.
\label{eqn:Stage1h*}
\eeq

Next, we determine the new force balance in the elastic drainage regime (which is distinctly different from the one found in stretching stage, equation~\eqref{eqn:Stage1FB}) via an `order analysis'~\cite{Chang1999} as follows. In the drainage stage, the filament has been stretched to a jet with a uniform radius $h^*$ of order $We^{-1/2}$ (equation~\eqref{eqn:Stage1h*}) and the elastic stresses are not negligible. `Polymer relaxation' must be included to effect the pull of the stretched polymers and counter the capillary driving force such that a slow drainage from the filament and into the beads can proceed. For relaxation to be included in the force balance, the terms $\frac{\partial \kappa}{\partial z}, \frac{1}{h^2} \left[ \frac{\partial (h^2 A_{zz})}{\partial z} \right]$ and $\frac{1}{h^2}\frac{\partial }{\partial z}\left[ h^2 \frac{\partial u}{\partial z} \right]$ must balance and will have the same order in equation~\eqref{eqn:SJmomentum}. A similar argument holds for the terms $\frac{d A_{zz}}{dt}, \left[ \frac{\partial u}{\partial z} \right] A_{zz}$ and the quadratic term, $\epsilon A^2_{zz}$, in equation~\eqref{eqn:SJAzz}; and for the terms $\frac{d A_{rr}}{dt}, \left[ \frac{\partial u}{\partial z} \right] A_{rr}$ and $\epsilon A_{rr} A_{zz}$ in equation~\eqref{eqn:SJArr}, respectively. For a BOAS configuration, the curvature, $\kappa$ (in equation~\eqref{eqn:SJmomentum}), varies from $h^{-1} \sim \mathcal{O}(We^{1/2})$ at the filament to $\mathcal{O}(1)$ at the bead~\cite{Chang1999}. We shall assign it the higher order (i.~e., $\mathcal{O}(We^{1/2})$) in our dominant force balance. Finally, the above information leads to the following scales for the variables involved in model~\eqref{eqn:SJmodel},
\beq
t \sim \mathcal{O}(We^{-1/2}), \,\, u \sim \mathcal{O}(We^{-1}), \,\, z \sim \mathcal{O}(We^{-3/2}), \,\, A_{zz} \sim \mathcal{O}(h^{-1}) = \mathcal{O}(We^{1/2}), \,\, A_{rr} \sim \mathcal{O}(We^{-1/2}),
\label{eqn:Scales}
\eeq
which leads to a new force balance at the leading order,
\beq
\frac{1}{Oh^2}\frac{\partial \kappa}{\partial z} + \frac{1-\nu}{h_0^2}\frac{\partial }{\partial z}\left[h_0^2 A_{zz_0}\right] + \frac{3\nu}{h_0^2}\frac{\partial }{\partial z} \left[h_0^2 \frac{\partial u_0}{\partial z}\right] = 0. \label{eqn:EDmomentum}
\eeq
Note that the order analysis in equation~\eqref{eqn:Scales}, especially for time, is different than the one obtained in the linear analysis by Chang~\cite{Chang1999}. This difference in the timescale is due to the impact of the finite stresses (see expression~\eqref{eqn:step5} for the exact form) obtained via the PTT stress constitutive relation. The kinematic condition~\eqref{eqn:Stage1kinematic} and the elastic stresses~(\ref{eqn:Stage1Arr}, \ref{eqn:Stage1Azz}) are governed by the ensuing, reduced set of equations along the characteristic lines, $u = \frac{d z}{d t}$,
\bseq \label{eqn:EDRegime}
\begin{align}
& \frac{d h_0^2}{d \delta} + h_0^2 \frac{\partial}{\partial z} (u_0) = 0, \label{eqn:EDkinematic} \\
& \frac{d A_{rr_0}}{d \delta} + A_{rr_0}\frac{\partial u_0}{\partial z} = -\frac{1}{We} \frac{\partial u_0}{\partial z} - \epsilon A_{rr_0} A_{zz_0}, \label{eqn:EDArr} \\
& \frac{d A_{zz_0}}{d \delta} -2 A_{rr_0}\frac{\partial u_0}{\partial z} = \frac{2}{We} \frac{\partial u_0}{\partial z} - \epsilon A^2_{zz_0}, \label{eqn:EDAzz}
\end{align}
\eseq
where $\delta = t / \sqrt{We}$. Note that, based on the order analysis,  we have approximated the quadratic terms in equation~(\ref{eqn:Stage1Arr}, \ref{eqn:Stage1Azz}) as follows: $(A_{rr_0} + A_{zz_0}) A_{rr_0} \approx A_{zz_0} A_{rr_0}$ and $(A_{rr_0} + A_{zz_0}) A_{zz_0} \approx A^2_{zz_0}$. We find a solution to the model~(\ref{eqn:EDmomentum}, \ref{eqn:EDRegime}) (unique unto a multiplicative constant) as follows: substituting a variable-separable form, $u_0(\delta, z) = \frac{U(\delta) f(z)}{We}, h_0(\delta, z) = h_1(\delta) h_2(z)$ in the kinematic condition~\eqref{eqn:EDkinematic}, leads to the following relation
\beq
\frac{2}{U} \frac{\dot{h_1}}{h_1} = - \left[ f' + 2f \frac{h'_2}{h_2} \right] = \text{constant},
\label{eqn:step1}
\eeq
where $(\,\dot{}\,), (\,\,)'$ denotes derivative with respect to $(\delta, z)$, respectively. In equation~\eqref{eqn:step1}, if we assume that the constant$=-1$, we arrive at an expression for $h_1(t)$,
\beq
h_1(\delta) = h_1(0) e^{-\int \frac{U(\delta)}{2} d\delta},
\label{eqn:step2}
\eeq
and, after utilizing the initial condition at the start of the elastic drainage stage (i.~e., $h(0) = h^*$) we find that $h_1(0)=h^*, h_2 = 1$ and $f(z) = z$ (using the second relation in equation~\eqref{eqn:step1}). At this stage we recall the experimental outcome of Entov and Hinch~\cite{Entov1997} demonstrating that the strain rate of a uniaxial extensional flow within the draining filament remains constant at two-thirds the rate at which the stress would relax at fixed strain, implying that we choose $U(t) = \nicefrac{2}{3}$, which leads us with the following set of solution
\bseq \label{eqn:step3}
\begin{align}
& h_0(t, z) = h^* e^{-\frac{t}{3\sqrt{We}}}, \label{eqn:step3a} \\
& u_0(t, z) = \frac{2 z}{3 We}. \label{eqn:step3b}
\end{align}
\eseq
A similar variable-separable form is assumed for the elastic stresses, leading us with the following set of differential equations,
\bseq \label{eqn:step4}
\begin{align}
& \frac{\dot{A_{rr_0}}}{A_{rr_0}} - 2\frac{\dot{h_1}}{h_1} + \epsilon A_{zz_0} = 0, \\
& \frac{\dot{A_{zz_0}}}{A_{zz_0}} + 4\frac{\dot{h_1}}{h_1} - \epsilon A_{zz_0} = 0,
\end{align}
\eseq
which is solved to arrive at the following expressions,
\bseq \label{eqn:step5}
\begin{align}
&{A_{zz_0}} = \dfrac{{A_{zz}}^*}{\dfrac{3 \epsilon}{4} + \exp{\left(\dfrac{-4 t / \sqrt{\text{We}}}{3}\right)}}, \label{eqn:step5a} \\
&{A_{rr_0}} = \dfrac{{A_{rr}}^* \exp{\left(-2t / \sqrt{\text{We}}\right)}}{\dfrac{3 \epsilon}{4} + \exp{\left(\dfrac{-4 t / \sqrt{\text{We}}}{3}\right)}}, \label{eqn:step5b}
\end{align}
\eseq
where ${A_{zz}}^* = \left[ {\left(\dfrac{1}{Oh^2 \left(1-\nu\right)}\right)}^{3/2} \right]\!\!\sqrt{We}$ and ${A_{rr}}^* = \dfrac{1}{{A_{zz}}^*}$.
The expressions given in relation~(\ref{eqn:step3}, \ref{eqn:step5}) represent the long-time asymptotic behavior of the system or the mean flow variables of the jet in a `quasi-stationary' configuration. In section \S \ref{sec:Results}, we linearize the slender jet model~\eqref{eqn:SJmodel} around the mean flow~(\ref{eqn:step3}, \ref{eqn:step5}) and perform a spatiotemporal stability analysis in order to identify the plausible flow-material parameter range leading to pinch-off.

\subsection{Linearization via normal mode expansion}\label{subsec:LA}
A major assumption of our analysis is that the evolution of the interface occurs on a slow timescale compared with the timescale of the perturbation (or the so called `quasi stationary' approximation). We restrict our stability analysis to the case when the disturbances are one-dimensional. In the ensuing description, we will denote the real/imaginary components with subscript $r/i$, respectively. Assuming an independent fate of each wavenumber, $\alpha$ (whose real part is chosen to be positive) and frequency, $\omega$, it is natural to consider disturbances in the form of a normal mode expansion, such that the axial velocity, the jet jet interface and the elastic stresses are expressed in terms of their mean values and perturbations as follows,
\beq
{\bf R} = {\bf R}_0 + {\bf R}' e^{i (\alpha z - \omega t)},
\label{eqn:Perturbation}
\eeq
%
where ${\bf R} = [u \,\, h \,\, A_{zz} \,\, A_{rr}]^T, {\bf R}_0 = [u_0 \,\, h_0 \,\, A_{zz_0} \,\, A_{rr_0}]^T, {\bf R}' = [A \,\, B \,\, C \,\, D]^T$ represents the total, the mean flow variables and the disturbance amplitudes, respectively. Next, substituting the solution form~\eqref{eqn:Perturbation} in the slender jet equations~\eqref{eqn:SJmodel} and retaining the terms which are linear in disturbance amplitude, we arrive at the equation governing the linearized force balance,
\begin{align}
&\left[ i h_0^2(\alpha u_0 - \omega -i \dfrac{\partial u_0}{\partial z} - i \alpha^2 \dfrac{3 \nu}{Re}) \right] A -\left[ i\alpha \dfrac{(1-\nu)}{Re} h_0^2 \right] (C - D) + \left[ 2 u_0 h_0 \dfrac{\partial u_0}{\partial z} - \dfrac{i \alpha}{Re Oh^2} \right.\nonumber\\
&\left. - \dfrac{h_0}{Re}\left[ 6\nu \dfrac{\partial^2 u_0}{\partial z^2} + i 6\nu \alpha \dfrac{\partial u_0}{\partial z} + 2i \alpha (1-\nu)(A_{zz_0}-A_{rr_0}) - \dfrac{i \alpha^3 h_0}{ Oh^2} \right] \right] B = 0,
\label{eqn:linearized1}
\end{align}
the linearized equation describing the kinematic equations,
\beq
\dfrac{i \alpha}{2}h_0^2 A + \left[ \dfrac{\partial h_0}{\partial t} -(i \omega)h_0 + h_0\dfrac{\partial u_0}{\partial z} + (i\alpha) u_0 h_0 \right] B = 0,
\label{eqn:linearized2}
\eeq
the axial component of the elastic stresses,
\beq
-2 i \alpha \left[ A_{zz_0} + \dfrac{1}{We} \right] A + \epsilon A_{zz_0} D + \left[ (i \alpha)u_0 - (i \omega) -2 \dfrac{\partial u_0}{\partial z} +\dfrac{1}{We} + \epsilon (2A_{zz_0}+A_{rr_0}) \right] C = 0,
\label{eqn:linearized4}
\eeq
and the radial component of elastic stresses,
\beq
i \alpha \left[ A_{rr_0} + \dfrac{1}{We} \right] A + \epsilon A_{rr_0} C + \left[ (i \alpha)u_0 - (i \omega) + \dfrac{\partial u_0}{\partial z} +\dfrac{1}{We} + \epsilon (A_{zz_0}+2A_{rr_0}) \right] D = 0.
\label{eqn:linearized3}
\eeq
Equations~(\ref{eqn:linearized1}-\ref{eqn:linearized4}) may be written in a matrix-vector format as follows,
\beq
\setlength{\arraycolsep}{4pt}
\renewcommand{\arraystretch}{1.0}
\left[
\begin{array}{cccc}
M^{11} & M^{12} & M^{13} & M^{14}\\
M^{21} & M^{22} & 0 & 0\\
M^{31} & 0 & M^{33} & M^{34}\\
M^{41} & 0 & M^{43} & M^{44}
\end{array}  \right] 
\left[
\begin{array}{c}
A \\
B \\
C \\
D \\
\end{array} \right]
= 0,
\label{eqn:DRP_matrix}
\eeq
where the expressions, $M^{ij}$, in the coefficient matrix are listed in \S \ref{sec:appendix}. A nontrivial solution for the system~\eqref{eqn:DRP_matrix}, imposes a zero determinant condition on the coefficient matrix which leads to the dispersion relation: $D(\alpha, \omega) = 0$ (see equation~\eqref{eqn:DRP} in \S \ref{sec:appendix}).

\subsection{Numerical Method}\label{subsec:NM}
The zeros of the dispersion relation ($D(\alpha, \omega)=0$, equation~\eqref{eqn:DRP}) were explored within the complex $\alpha-\omega$ plane inside the region $... \le \omega_r \le ..., 0.0 \le \omega_i \le ..., \alpha_r \le ... $ and $|\alpha_i| \le ...$. The first step in the spatiotemporal analysis entails the procedure of finding the most unstable mode for real wavenumber, $\alpha$. This mode is the largest positive imaginary component of any root of the dispersion relation, also known as the temporal growth rate, $\omega_{\text{Temp}}$~\cite{Sircar2019}. In other words, this step consists of detecting the admissible saddle points ($\omega \in \mathbb{C}, \alpha \in \mathbb{R}$) satisfying the equations~\cite{Huerre1990},
\bseq \label{eqn:wTemp}
\begin{align}
&D(\alpha, \omega) = 0, \label{eqn:wTemp1} \\
&\dfrac{\partial \omega_i}{\partial \alpha} = \dfrac{\nicefrac{\partial D}{\partial \alpha}}{\nicefrac{\partial D}{\partial \omega_i}} = 0, \label{eqn:wTemp2}
\end{align}
\eseq
and then among all the possible roots of equation~\eqref{eqn:wTemp}, we identify those roots with the largest positive imaginary component of the frequency. Equation~\eqref{eqn:wTemp} (refer equations~(\ref{eqn:DRP}-\ref{eqn:DRP_imag}), in \S \ref{sec:appendix} for the detailed expressions) is solved using a multivariate Newton-Raphson algorithm. 

Next, the eigenpairs with complex wavenumbers and frequencies are permitted in the solution of equation~\eqref{eqn:wTemp} and the absolute growth rate (or the growth rate at the cusp point, $\omega^{\text{cusp}}_i$) is numerically evaluated, starting from the temporal growth rate, $\omega^{\text{Temp}}_i$. The cusp point in the $\omega-$plane is a saddle point satisfying the criteria: $D(\alpha, \omega^{\text cusp})\!=\!\frac{\partial D(\alpha, \omega^{\text cusp})}{\partial \alpha}\!=\!0$ but $\frac{\partial^2 D(\alpha, \omega^{\text cusp})}{\partial \alpha^2}\!\ne \!0$. However, not all cusp points are unstable and, in particular, the evanescent modes (time-asymptotically stable or false modes) are segregated from the regular cusp points using the Briggs idea of analytic continuation, as follows~\cite{Kupfer1987}. In the $\omega$-plane, we draw a ray parallel to the $\omega_i$-axis from the cusp point such that it intersects the $\alpha_i = 0$ curve and count the number of intersections. If the ray drawn from the cusp point intersects the $\alpha_i=0$ curve even number of times, then the flow dynamics correspond to an evanescent mode. Otherwise, in the case of odd intersections the observed cusp point is genuine, leading to either absolutely unstable system (in the upper half of the $\omega$-plane) or convectively unstable system (in the lower half of the $\omega$-plane); provided the system is temporally unstable~\cite{Bansal2021}.

\section{Results}\label{sec:Results}
The numerical continuation of the points on the absolute growth rate curves was achieved in the strong extensional regime (i.~e., within the range $We \in [100, 400]$) and two specific values of $\nu$: $\nu=0.01$ (or the elastic stress dominated case) and $\nu=0.99$ (or the viscous stress dominated case, refer \S \ref{subsubsec:VE}), using a discrete step-size of $\triangle We = 10^{-3}$. In the limit $We \ll 1$, the breakup dynamics is predominantly governed by the elasto-capilliary forces~\cite{Clasen2006}. The viscous as well as the inertial forces are relatively small and for this reason the Reynolds number was fixed in our studies at $Re=10$ (the roots of the dispersion relation~\eqref{eqn:DRP} were found to be practically invariant when the Reynolds number was varied within the range, $0.1 \le Re \le 100$). Further four values of the Ohnesorge number, $Oh = 0.00018, 0.1, 0.25, 0.5$ were chosen, based on the experimental outcome on the dynamics of bead formation~\cite{Bhat2010} (suggesting an appropriate range of $0 \le Oh \le 1$, for the jetting regime). Finally, the flow in the elastic drainage regime are compared at two values of the maximum elongation parameter: $\epsilon=0.01$ (the weak extension case) and $\epsilon=0.99$ (the strong extension case, refer \S \ref{subsubsec:FE}).

The parameter, $t$, in expressions~(\ref{eqn:DRP}-\ref{eqn:DRP_imag}), representing the time of evolution of the jet interface (a slow manifold) is fixed at $t=0.001$ and $t=10.0$. Changes in this parameter beyond $t=10.0$ (i.~e., within the range from $10.0 \le t \le 50.0$) induces a change of less than $1\%$ in the eigenvalues (or the roots of the dispersion relation). The instability studies are presented at two spatial locations in the axial direction, (a) $z=0.05$ or near jet exit ($z=0.0$ is the jet exit where the solution is known via the boundary conditions), and (b) $z=20.0$ or further downstream (see Fig.~\ref{fig:Fig2}-\ref{fig:Fig5}). 

%

\subsection{Spatiotemporal stability analysis}\label{subsec:STI}
The Oldroyd-B liquid jet ($\epsilon=0$, equation~\eqref{eqn:EStress}) allows the polymer chains (modeled as linear dumbbells) to be stretched infinitely leading to an issue of predicting infinite stress at finite strain rates~\cite{Chang1999}. As a result, iterated recoil and stretching dynamics can proceed indefinitely in the filaments of Oldroyd-B jets. Conversely, when finite extensibility comes into play (via the PTT model), the drainage is relatively rapid for the recoil instability to take effect and instead one observes pinch-off at finite times. The evolution of the minimum jet radius, $h_{\text{min}}$ (since the jet tapers along the axial direction, the minimum radius is found at the maximum axial distance, $z_m$, beyond which the dispersion relation (equation~\eqref{eqn:DRP}) has no solution within the search region) for the PTT fluid at fixed parameter values, $Oh=0.5, We=100$ and $Re=10$ are shown in Fig.~\ref{fig:Fig6}. Notice the presence of two distinct stages prior to pinch-off. The stretching stage (the elastic drainage stage) characterized by a precipitous (relatively slow) drop in the jet radius, respectively. The presence of finite stresses ($\epsilon \ne 0$, equation~\eqref{eqn:step5}) at constant strain rate (equation~\eqref{eqn:step3b}) implies a fixed cut-off time at which the finite elongation effects set in. The spatiotemporal analysis in \S \ref{subsubsec:TS} \ref{subsubsec:VE} \ref{subsubsec:FE} is performed before this cut-off time.
\bef
\centering
\includegraphics[width=0.55\linewidth, height=0.4\linewidth]{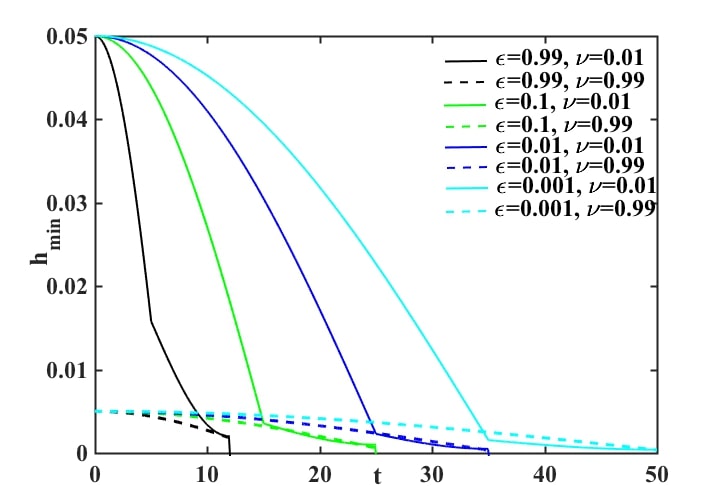}
\caption{The minimum radius of the jet, $h_{\text{min}}=h(z_m)$ (where $z_m$ is the maximum axial distance beyond which the dispersion relation (equation~\eqref{eqn:DRP}) has no solution) versus time, $t$, and at fixed values of the flow parameters, $Oh=0.5, We=100$ and $Re=10$.}
\label{fig:Fig6}
\eef

Spatiotemporal analysis is typically relevant when one introduces an impulse excitation locally in a flow and observes how that disturbance evolves with time. In an effort to determine the range of $We$ and $Oh$ (for fixed $Re$ and $\nu$) in which the flow regimes are absolutely unstable, convectively unstable or temporally stable, we recover the absolute growth rates~(Figures.~\ref{fig:Fig2},~\ref{fig:Fig3},~\ref{fig:Fig4}). A discontinuity or an absence of the absolute growth rate curves Figures.~\ref{fig:Fig2},~\ref{fig:Fig3},~\ref{fig:Fig4} indicates a region of temporal stability. 

\subsubsection{Impact of temporal stretching}\label{subsubsec:TS}
\begin{figure}[htbp]
\centering
\begin{subfigure}{0.48\textwidth}
\includegraphics[width=0.99\linewidth, height=0.8\linewidth]{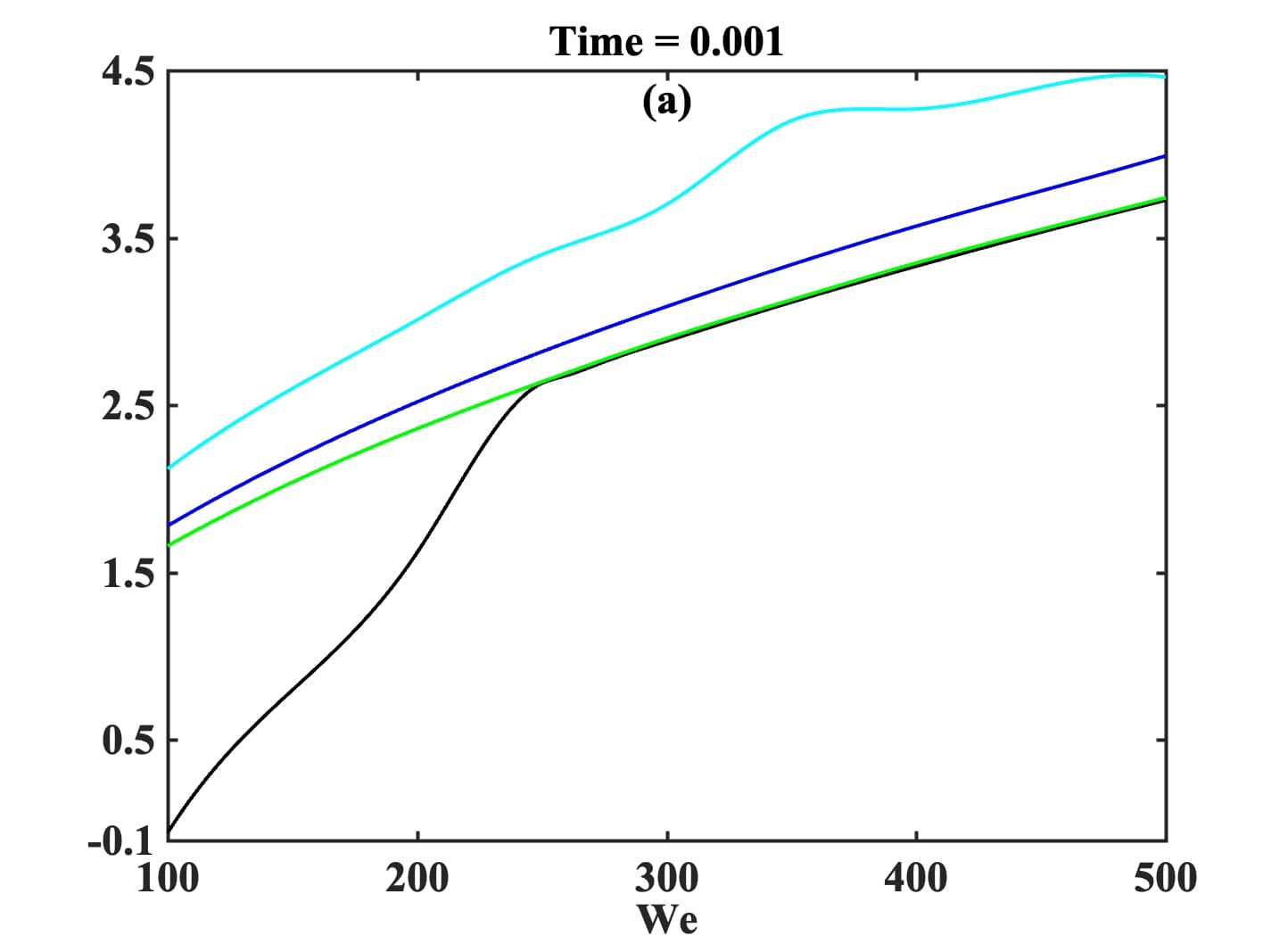}
\vskip -17pt
\caption*{} 
\end{subfigure}
\begin{subfigure}{0.48\textwidth}
 \includegraphics[width=0.99\linewidth, height=0.8\linewidth]{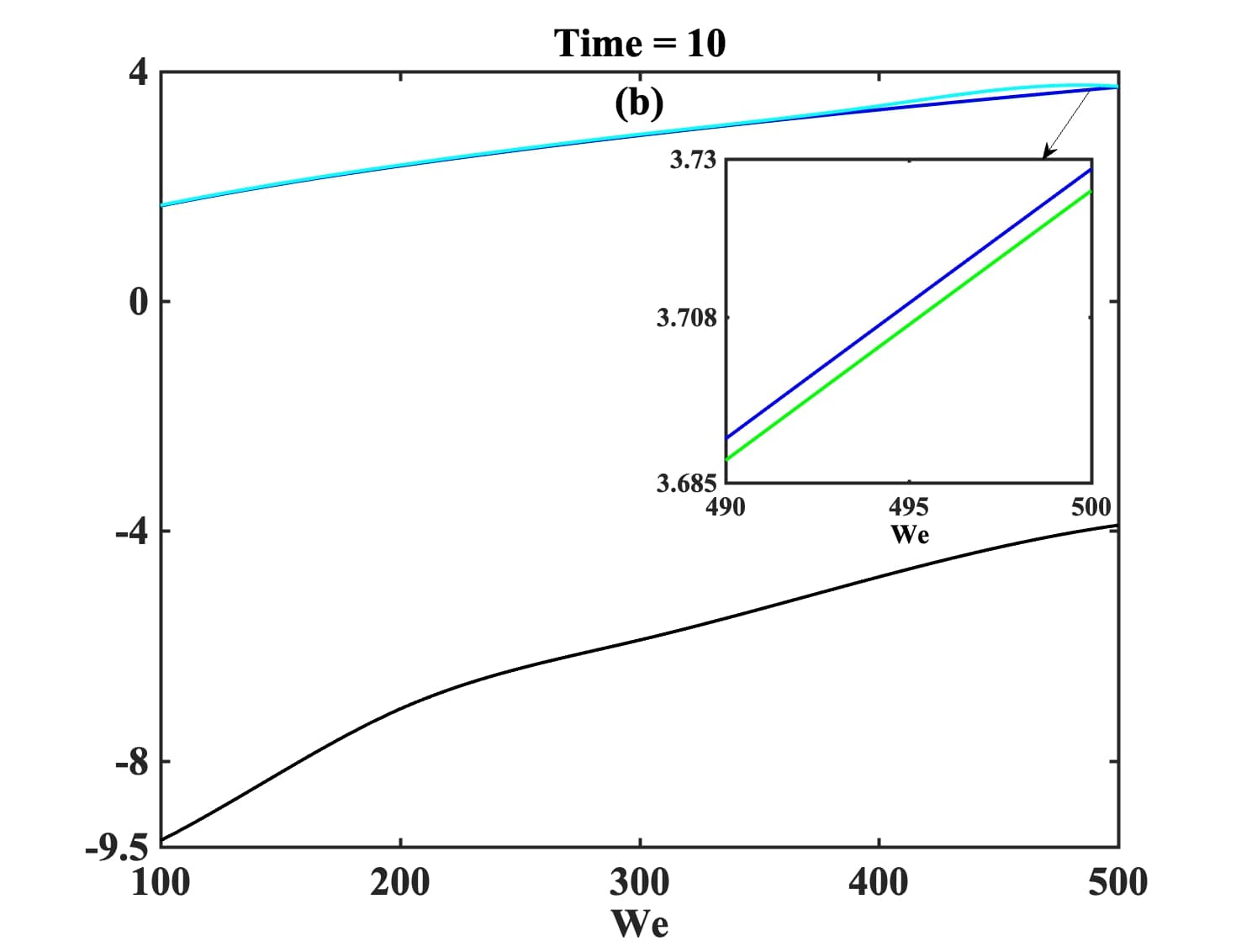}
 \vskip -17pt
 \caption*{} 
\end{subfigure}
\begin{subfigure}{0.48\textwidth}
\includegraphics[width=0.99\linewidth, height=0.8\linewidth]{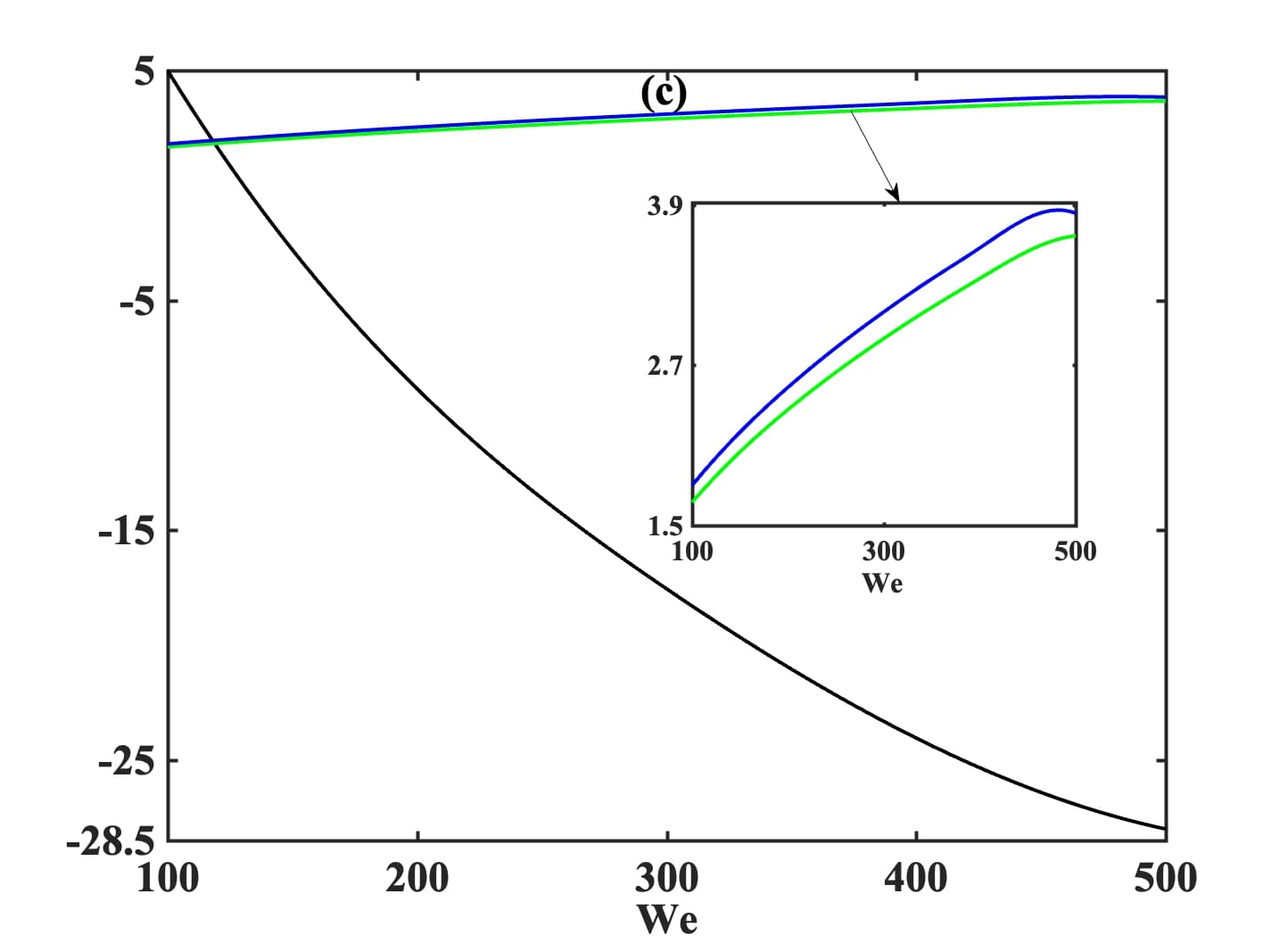}
\vskip -17pt
\caption*{} 
\end{subfigure}
\begin{subfigure}{0.48\textwidth}
 \includegraphics[width=0.99\linewidth, height=0.8\linewidth]{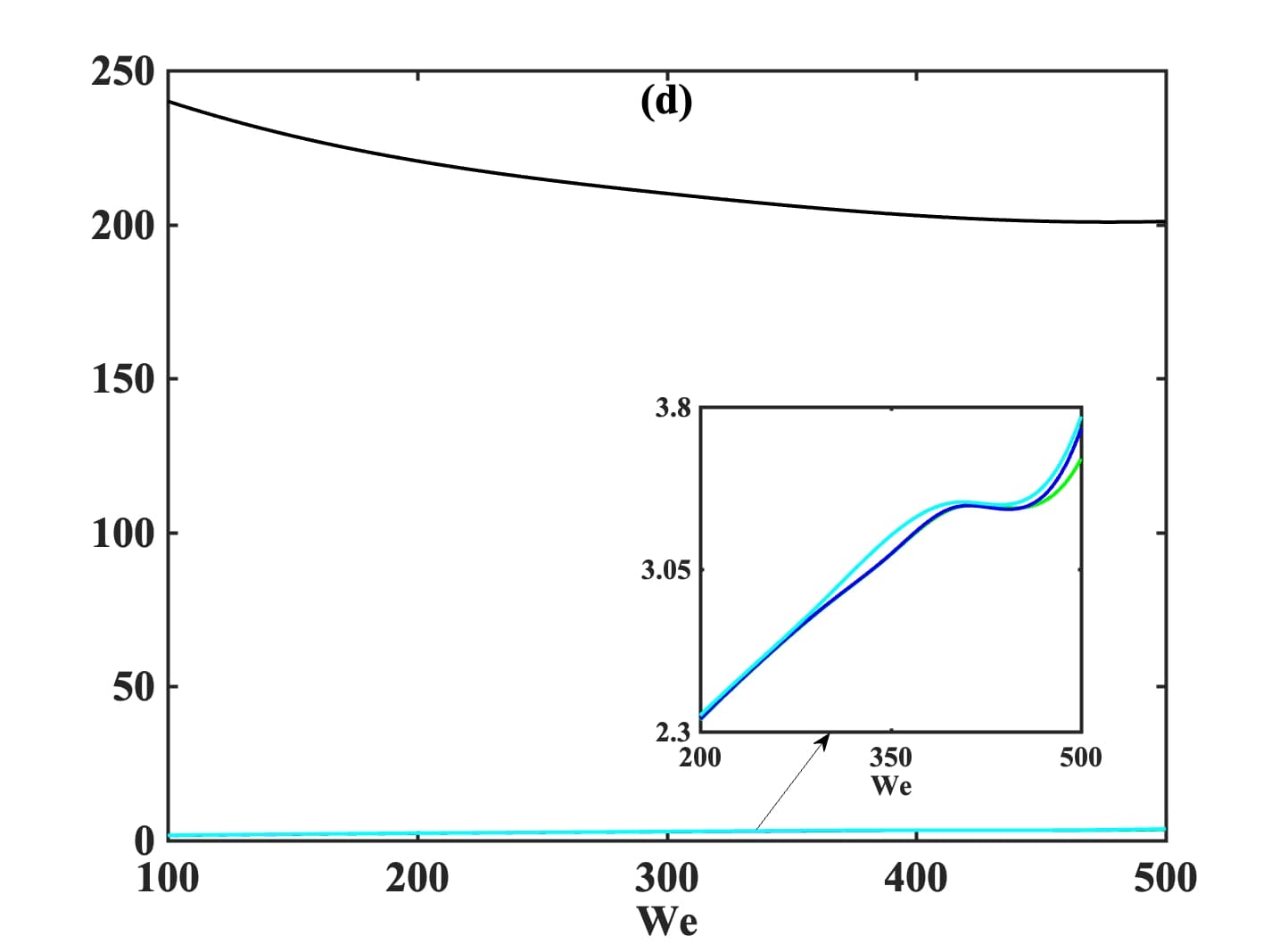}
 \vskip -17pt
 \caption*{} 
\end{subfigure}
\caption{The cusp point, $\omega^{\text{cusp}}_i$ versus $We$ evaluated at viscosity ratio, $\nu=0.01$, elongation parameter, $\epsilon=0.01$, and at spatiotemporal location (a) $t=0.001, z=0.05$, (b) $t=10.0, z=0.05$, (c) $t=0.001, z=20.0$, and (d) $t=10.0, z=20.0$, shown at Ohnesorge numbers $Oh=0.00018$ (\protect\blackline), $Oh=0.1$ (\protect\greenline), $Oh=0.25$ (\protect\blueline) and $Oh=0.5$ (\protect\cyanline), respectively.}
\label{fig:Fig2}
\end{figure}

The jetting/breakup process in the elastic drainage stage, involves a delicate interplay of capillary/surface tension forces (which retains the shape of the jet) and the elastic forces (which tends to elongate the liquid thread) leading to the formation of a quasi-stationary BOAS structure, whose spatiotemporal stability (via the absolute growth rate curves) with changing time of evolution, $t$, in the $Oh-We$ parameter space (at $\nu=0.01$ and $\epsilon=0.01$) are presented in Fig.~\ref{fig:Fig2}. The stabilizing effects of the capillary forces near the nozzle ($z=0.05$, Fig.~\ref{fig:Fig2}a,b) are visible at infinitesimally low $Oh$ values (i.~e., at $Oh = 0.00018$. Ohnesorge number $\propto (\text{capillary forces})^{-1/2}$), where we find that the disturbance is convectively unstable within a partial range at early times (i.~e., within the range $100 \le We \le 110$, Fig~\ref{fig:Fig2}a) and within the entire range of $We$ values at later times. A similar finding (of convective instability at very low values of $Oh$) is reported further downstream (Fig~\ref{fig:Fig2}c). The $Oh = 0.00018$ curve at spatiotemporal location, $t = 10.0, z = 20.0$ (Fig.~\ref{fig:Fig2}d), represents an evanescent, or time-asymptotically stable mode (refer phase diagram, Fig.~\ref{fig:Fig5}d). Secondly note, that the absolute growth rates, at identical $Oh$ and $We$ values, are lower at later times (i.~e., comparing the left versus the right column in Fig.~\ref{fig:Fig2}, and ignoring the exceptional case of the $Oh = 0.00018$ curve in Fig.~\ref{fig:Fig2}d, as explained above). This observation is attributed to an increase in the axial stress (refer equation~\eqref{eqn:step5a}) at constant strain rate (equation~\eqref{eqn:step3b}), at asymptotically long times: the so called `strain-hardening' phenomena. Experiments~\cite{Christanti2002} have shown that viscoelastic liquids which exhibit strain-hardening are relatively stable with longer breakup lengths. Finally, notice the presence of temporal stability at early times and near the nozzle (i.~e., $Oh = 0.5$ curve in Fig.~\ref{fig:Fig2}c is temporally stable. More details provided in the phase diagram, Fig.~\ref{fig:Fig5}c), in sync with the experimental findings by McKinley~\cite{Bhat2010} who demonstrated that both the bead formation as well as the jet breakup is suppressed at low surface tension (or large values of $Oh$).

\subsubsection{Impact of viscoelasticity}\label{subsubsec:VE}
\begin{figure}[htbp]
\centering
\begin{subfigure}{0.48\textwidth}
\includegraphics[width=0.99\linewidth, height=0.8\linewidth]{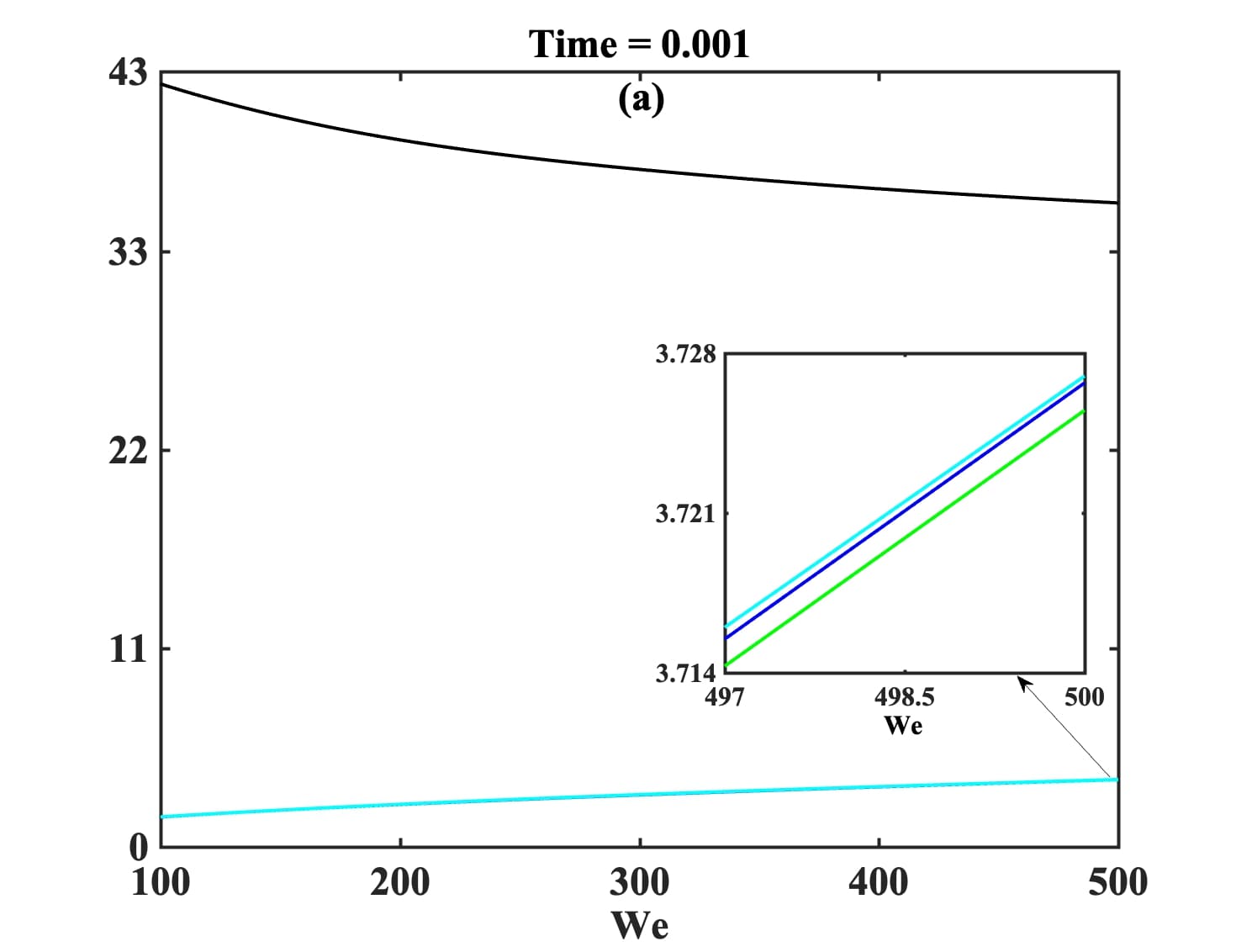}
\vskip -17pt
\caption*{} 
\end{subfigure}
\begin{subfigure}{0.48\textwidth}
 \includegraphics[width=0.99\linewidth, height=0.8\linewidth]{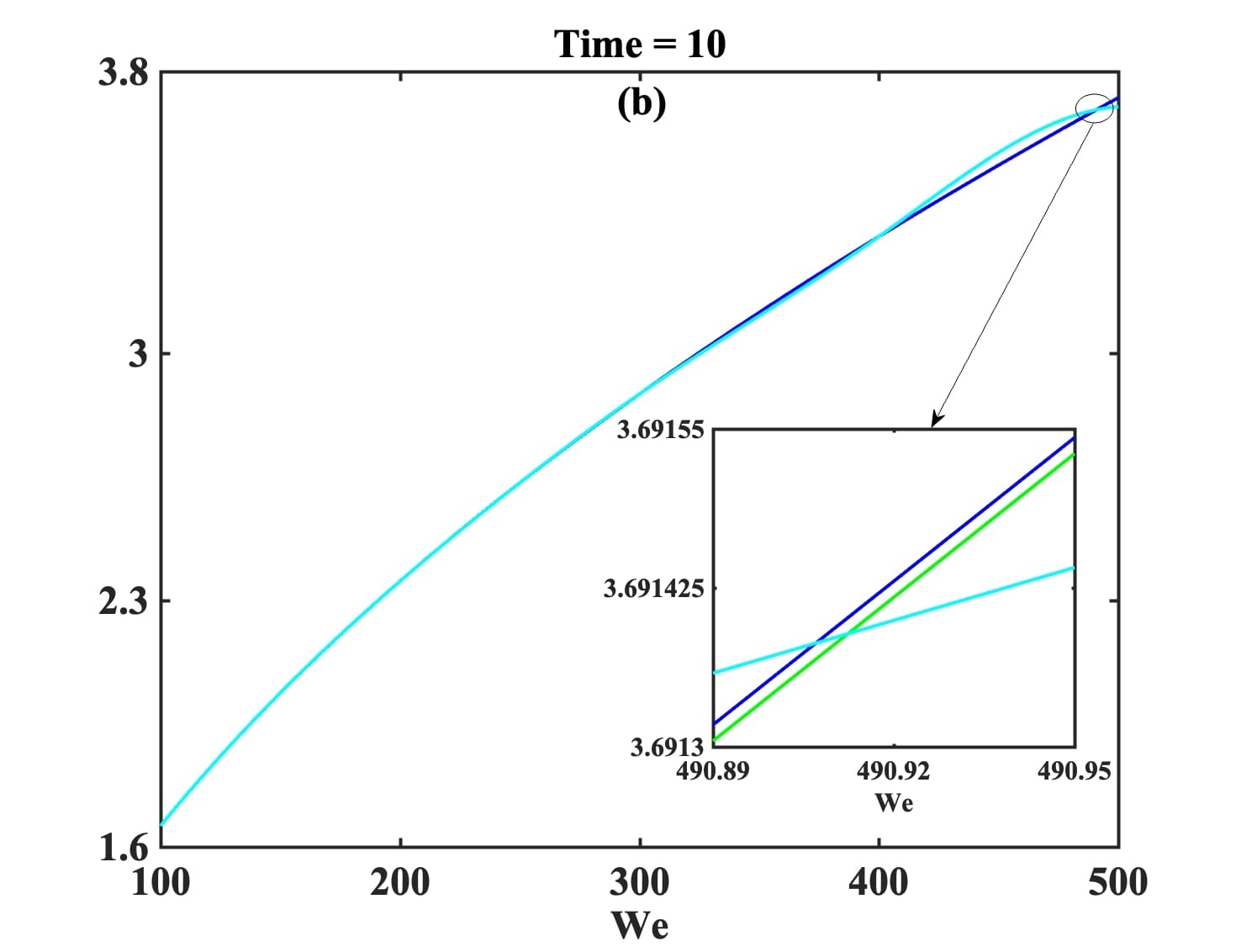}
 \vskip -17pt
 \caption*{} 
\end{subfigure}
\begin{subfigure}{0.48\textwidth}
\includegraphics[width=0.99\linewidth, height=0.8\linewidth]{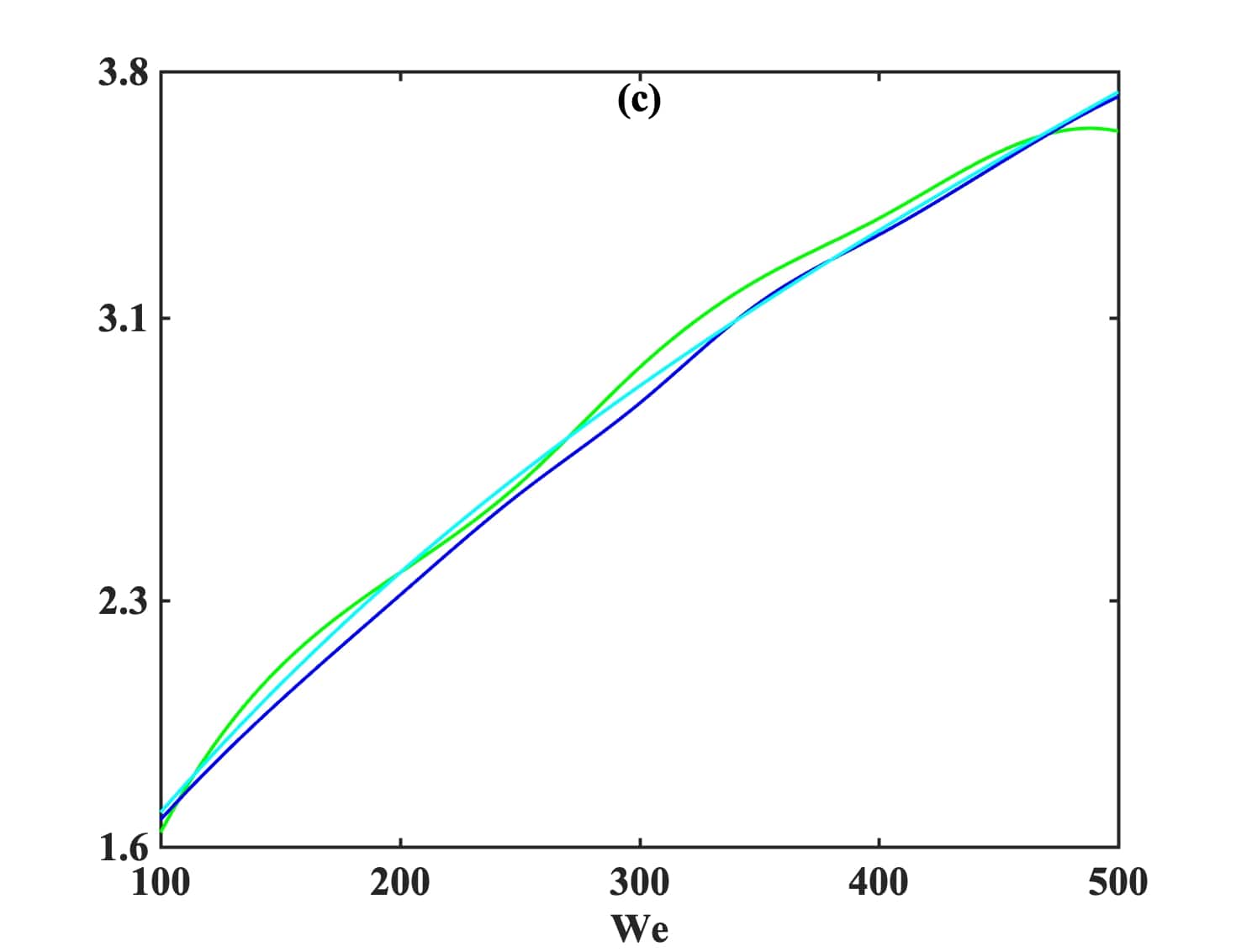}
\vskip -17pt
\caption*{} 
\end{subfigure}
\begin{subfigure}{0.48\textwidth}
 \includegraphics[width=0.99\linewidth, height=0.8\linewidth]{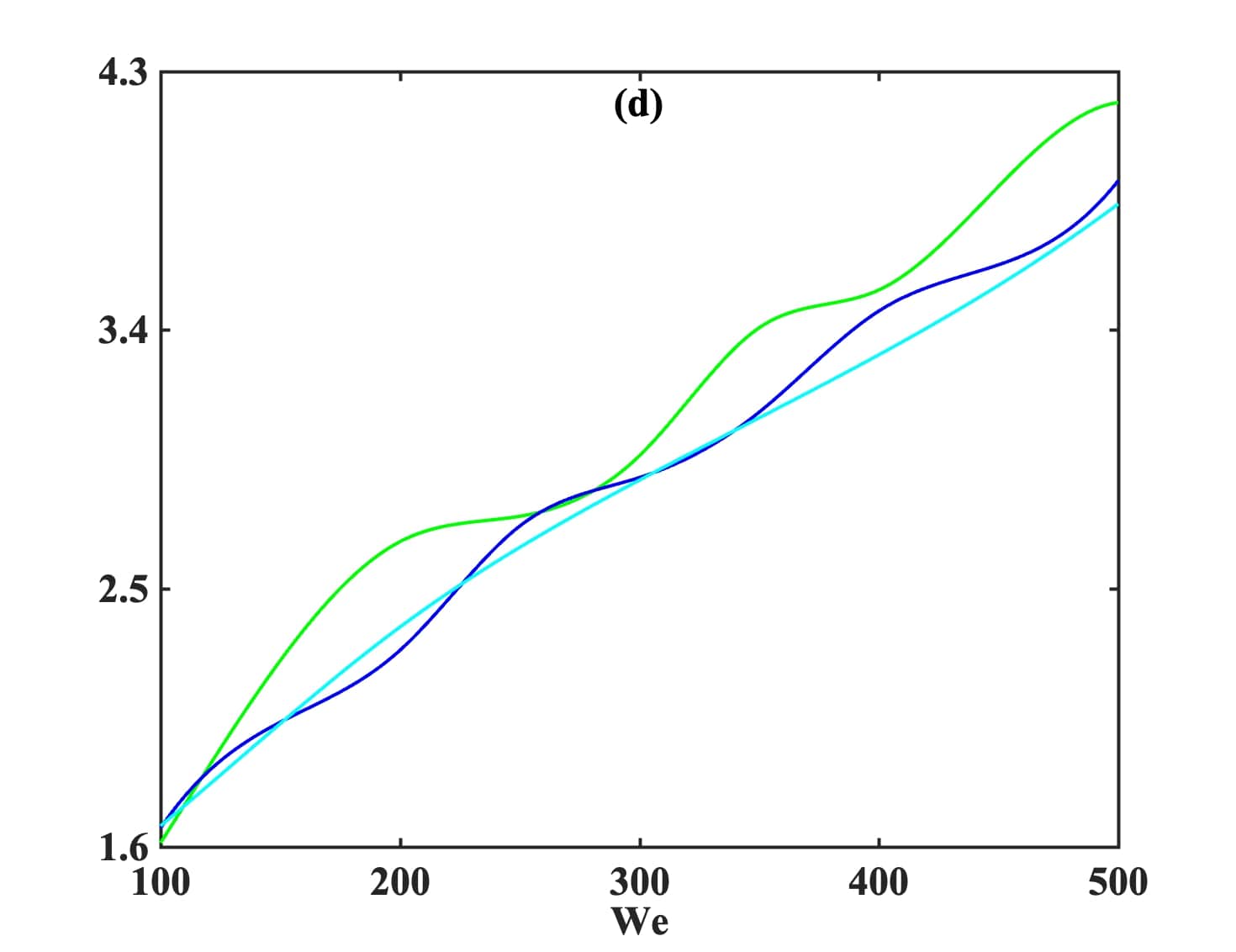}
 \vskip -17pt
 \caption*{} 
\end{subfigure}
\caption{The cusp point, $\omega^{\text{cusp}}_i$ versus $We$ evaluated at viscosity ratio, $\nu=0.99$, elongation parameter, $\epsilon=0.01$, and at spatiotemporal location (a) $t=0.001, z=0.05$, (b) $t=10.0, z=0.05$, (c) $t=0.001, z=20.0$, and (d) $t=10.0, z=20.0$, shown at Ohnesorge numbers $Oh=0.00018$ (\protect\blackline), $Oh=0.1$ (\protect\greenline), $Oh=0.25$ (\protect\blueline) and $Oh=0.5$ (\protect\cyanline), respectively.}
\label{fig:Fig3}
\end{figure}

The elasto-capilliary balance of the absolute growth rate curves in the $Oh-We$ parameter space, for a elastic-stress dominated dominated case ($\nu=0.99$), is shown in Fig.~\ref{fig:Fig3}. The stabilizing influence of capillary forces are again visible at low values of $Oh$, e.~g., notice the presence of either an evanescent mode (Fig.~\ref{fig:Fig3}a with corresponding phase diagram shown in Fig.~\ref{fig:Fig5}e) or temporal stability (Fig.~\ref{fig:Fig3}b-d with corresponding phase diagram shown in Fig.~\ref{fig:Fig5}f-h, respectively) at $Oh=0.00018$ and within the entire range of $We$ values under investigation. Further, notice that elasticity has a stabilizing impact at later times and at larger values of $We$ (e.~g., the absolute growth rate at $Oh=0.5$ is marginally lower than the absolute growth rate at $Oh=0.1, 0.25$, within the range $We \ge 490$, Fig.~\ref{fig:Fig3}b, d): a finding which is coetaneous with the theoretical studies of Patrascu~\cite{Patrascu2018} which reported that the BOAS structure is preserved and the final breakup is retarded by the buildup of large extensional stresses at high $We$ values.

\subsubsection{Impact of finite extensibility}\label{subsubsec:FE}
\begin{figure}[htbp]
\centering
\begin{subfigure}{0.48\textwidth}
\includegraphics[width=0.99\linewidth, height=0.8\linewidth]{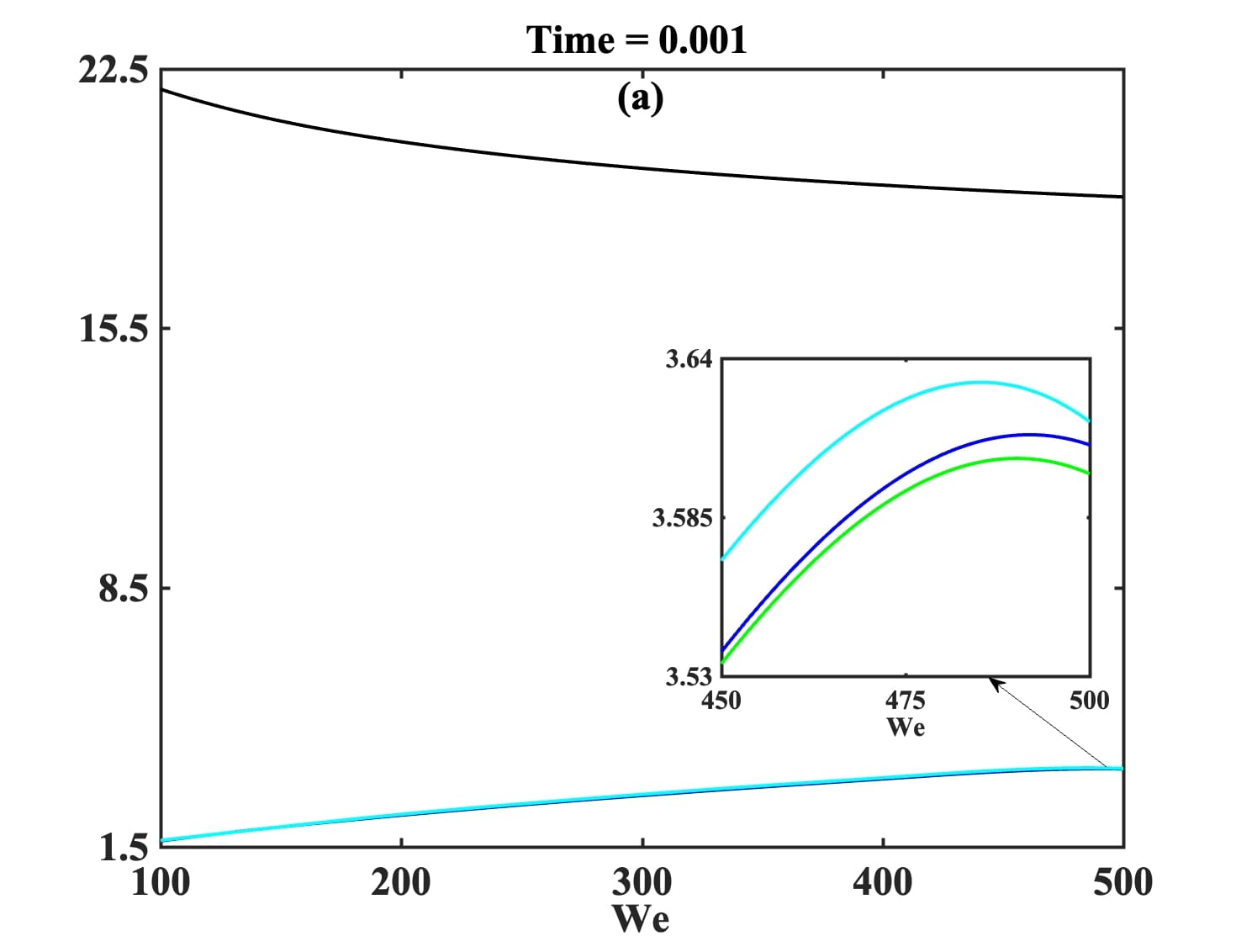}
\vskip -17pt
\caption*{} 
\end{subfigure}
\begin{subfigure}{0.48\textwidth}
 \includegraphics[width=0.99\linewidth, height=0.8\linewidth]{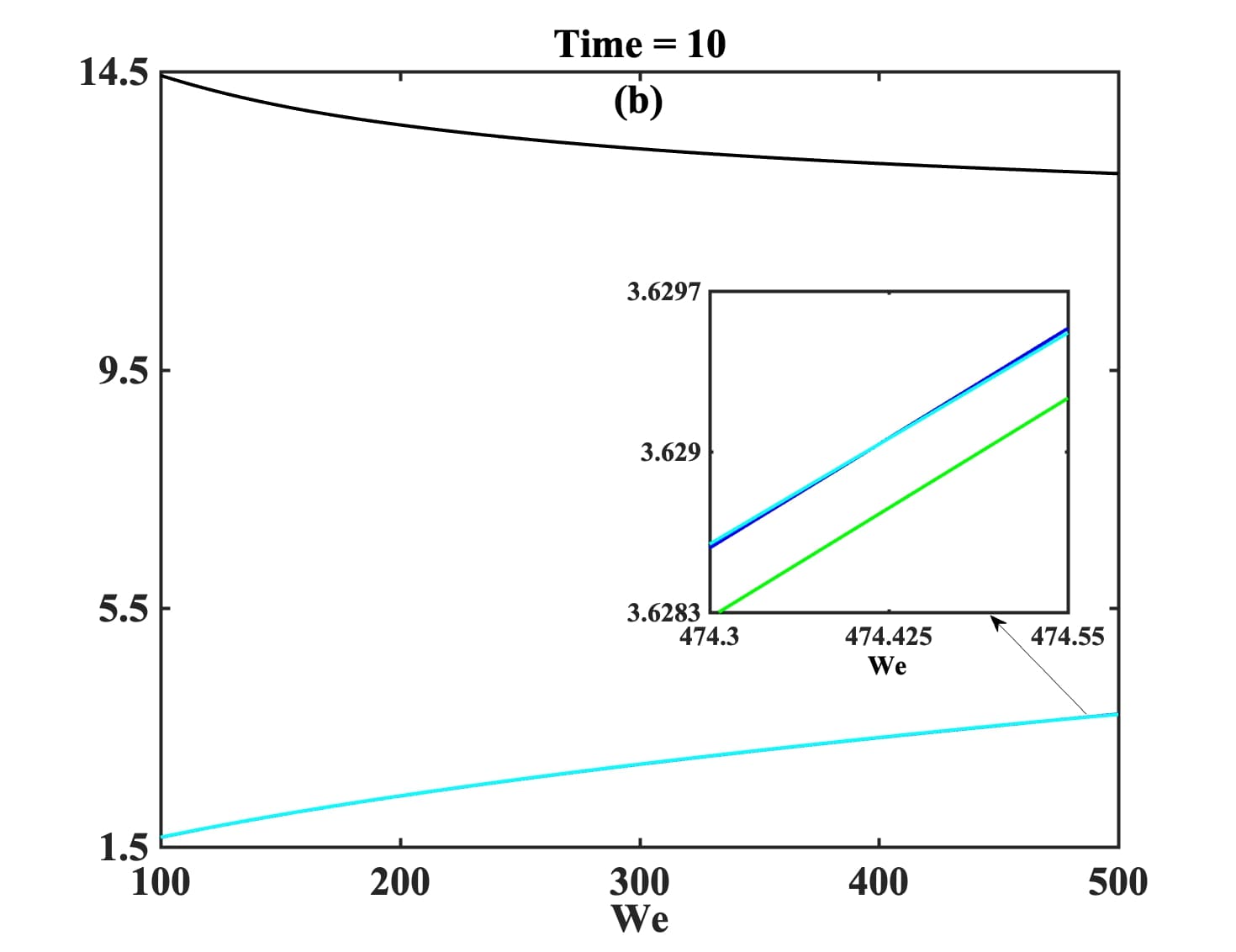}
 \vskip -17pt
 \caption*{} 
\end{subfigure}
\begin{subfigure}{0.48\textwidth}
\includegraphics[width=0.99\linewidth, height=0.8\linewidth]{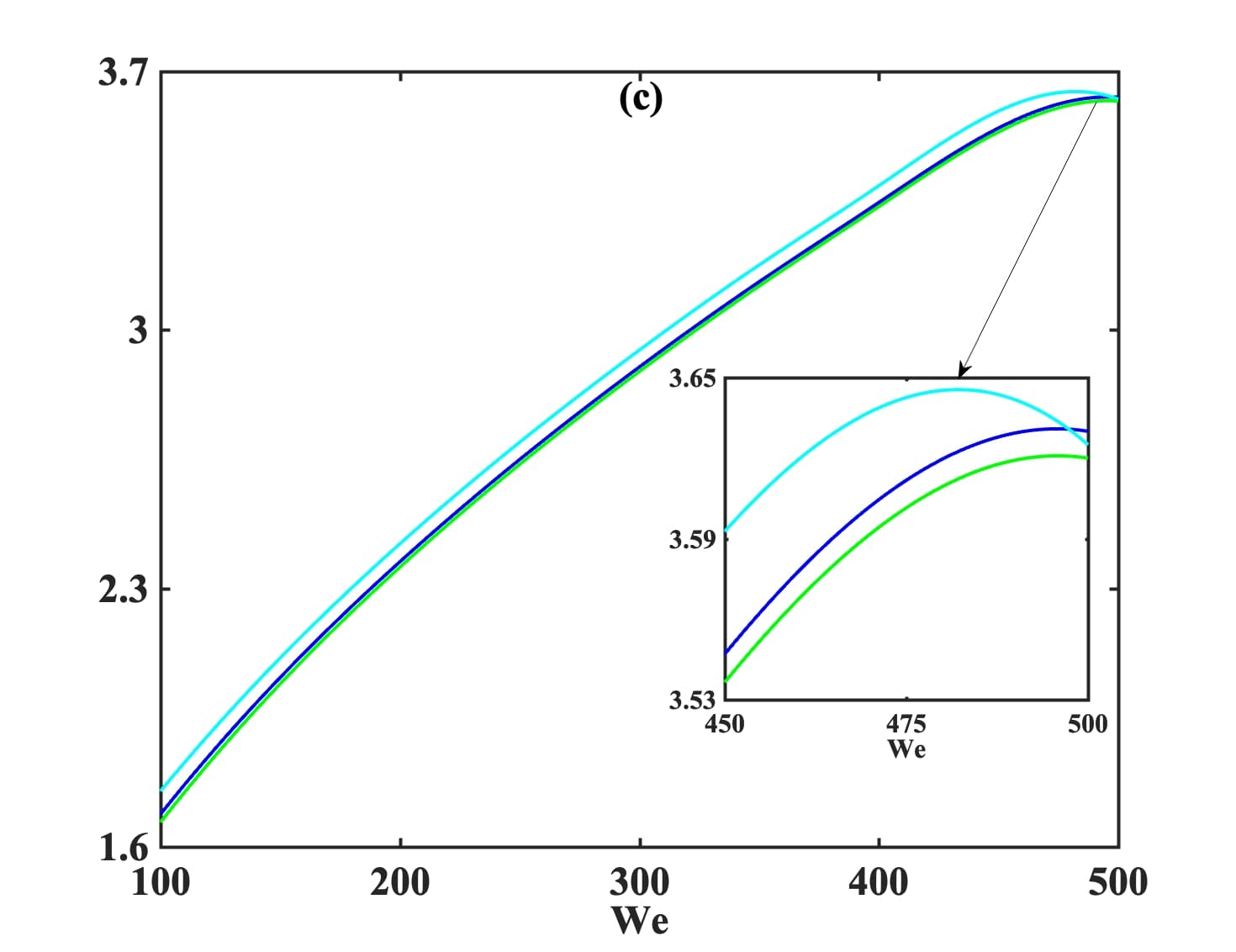}
\vskip -17pt
\caption*{} 
\end{subfigure}
\begin{subfigure}{0.48\textwidth}
 \includegraphics[width=0.99\linewidth, height=0.8\linewidth]{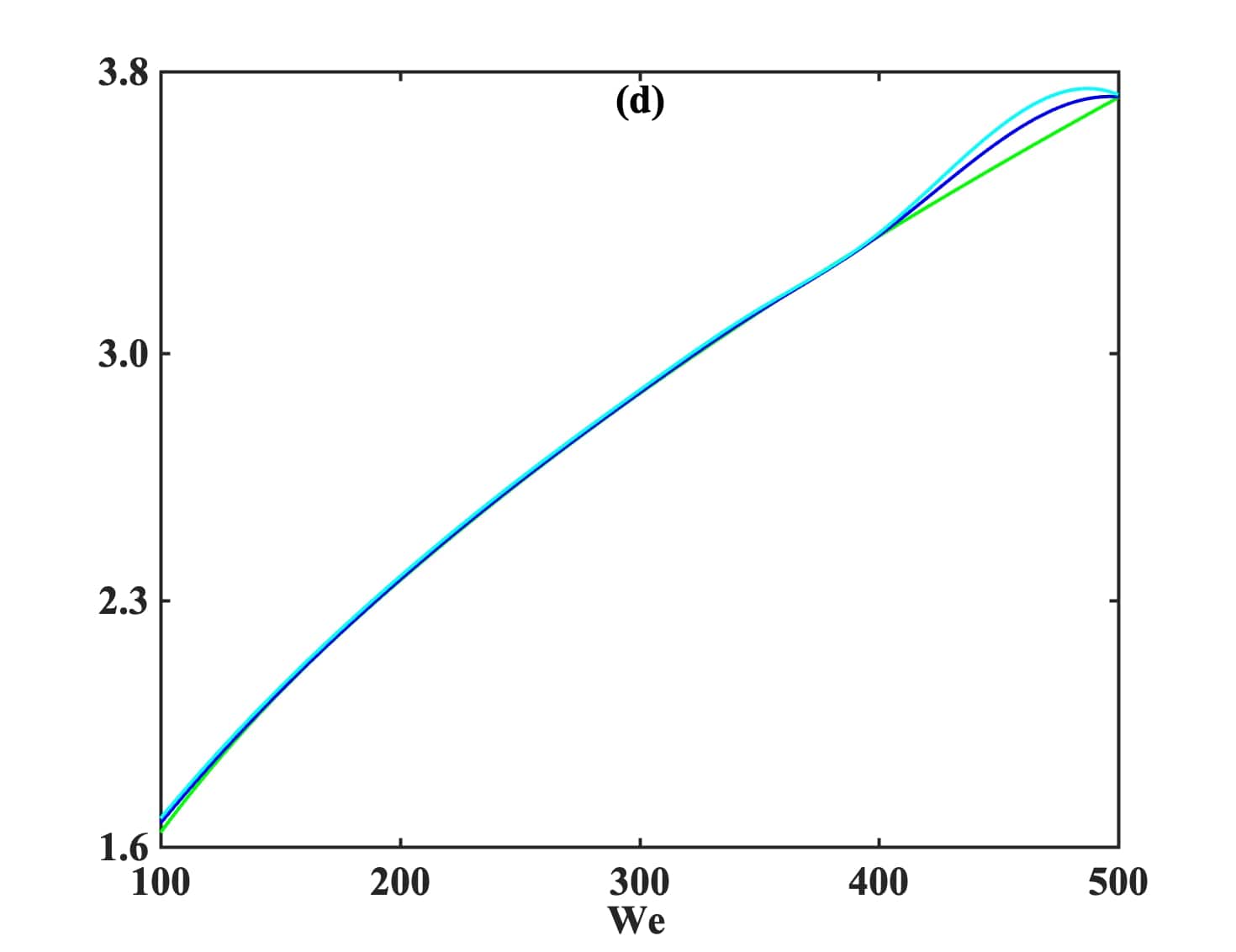}
 \vskip -17pt
 \caption*{} 
\end{subfigure}
\caption{The cusp point, $\omega^{\text{cusp}}_i$ versus $We$ evaluated at viscosity ratio, $\nu=0.01$, elongation parameter, $\epsilon=0.99$, and at spatiotemporal location (a) $t=0.001, z=0.05$, (b) $t=10.0, z=0.05$, (c) $t=0.001, z=20.0$, and (d) $t=10.0, z=20.0$, shown at Ohnesorge numbers $Oh=0.00018$ (\protect\blackline), $Oh=0.1$ (\protect\greenline), $Oh=0.25$ (\protect\blueline) and $Oh=0.5$ (\protect\cyanline), respectively.}
\label{fig:Fig4}
\end{figure}

Finally, the effect of the maximum elongation parameter, $\epsilon$, on the evolution of the absolute growth rate curves in the $Oh-We$ parameter space, are depicted in Fig~\ref{fig:Fig4}. Comparing the growth rate curves in Fig~\ref{fig:Fig4} (computed at $\epsilon=0.99$ and $\nu=0.01$) with the corresponding curves in Fig~\ref{fig:Fig2} (evaluated at $\epsilon=0.01$ and $\nu=0.01$), the following strain-hardening impact on the growth rates is visible: the absolute growth rates at lower value of $\epsilon$, is lower. This conclusion is again ascribed due to an increasing axial stress (equation~\eqref{eqn:step5a}) via lowering the elongation parameter, $\epsilon$.

\begin{figure}[htbp]
\centering
\begin{subfigure}{0.245\textwidth}
\includegraphics[width=1.1\linewidth, height=0.9\linewidth]{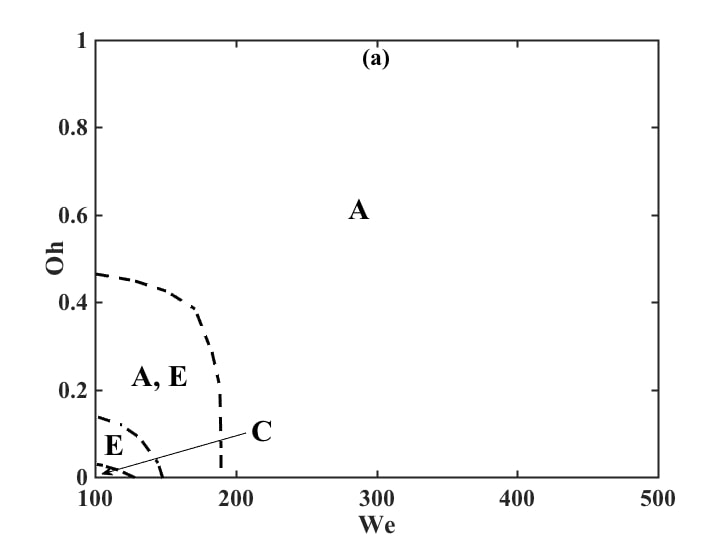}
\vskip -17pt
\caption*{} 
\end{subfigure}
\begin{subfigure}{0.245\textwidth}
 \includegraphics[width=1.1\linewidth, height=0.9\linewidth]{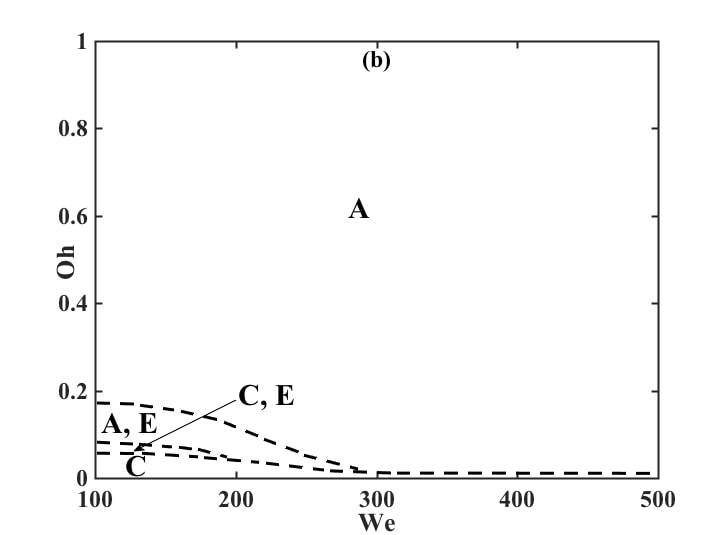}
 \vskip -17pt
 \caption*{} 
\end{subfigure}
\begin{subfigure}{0.245\textwidth}
\includegraphics[width=1.1\linewidth, height=0.9\linewidth]{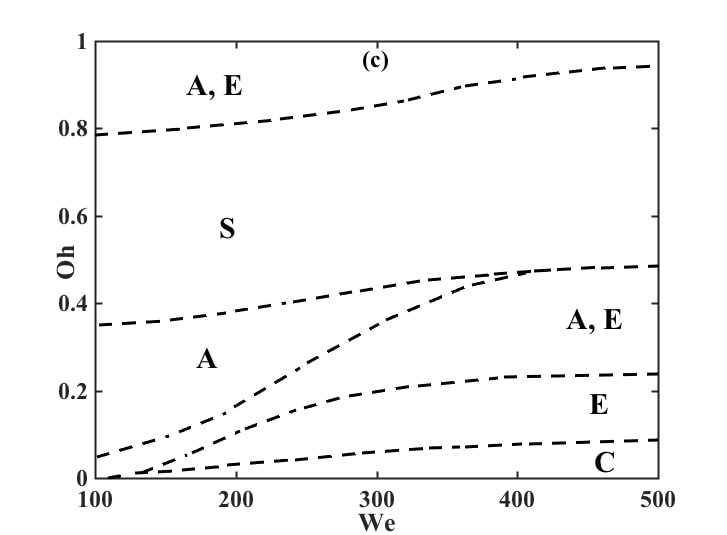}
\vskip -17pt
\caption*{} 
\end{subfigure}
\begin{subfigure}{0.245\textwidth}
 \includegraphics[width=1.1\linewidth, height=0.9\linewidth]{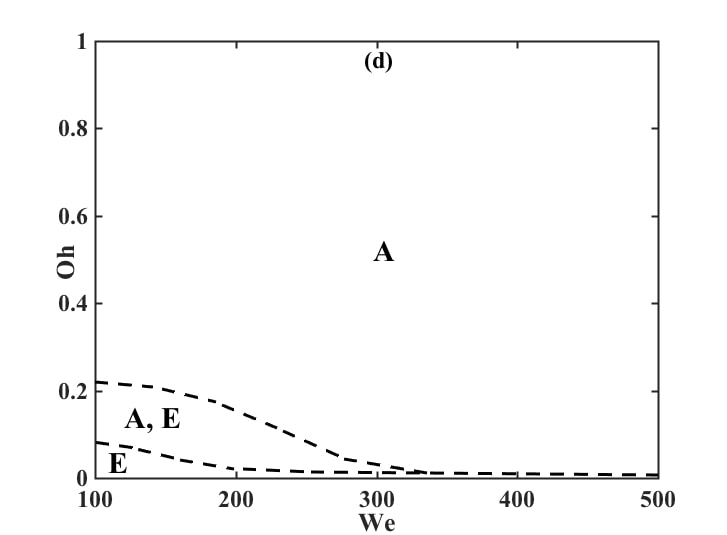}
 \vskip -17pt
 \caption*{} 
\end{subfigure}
\begin{subfigure}{0.245\textwidth}
\includegraphics[width=1.1\linewidth, height=0.9\linewidth]{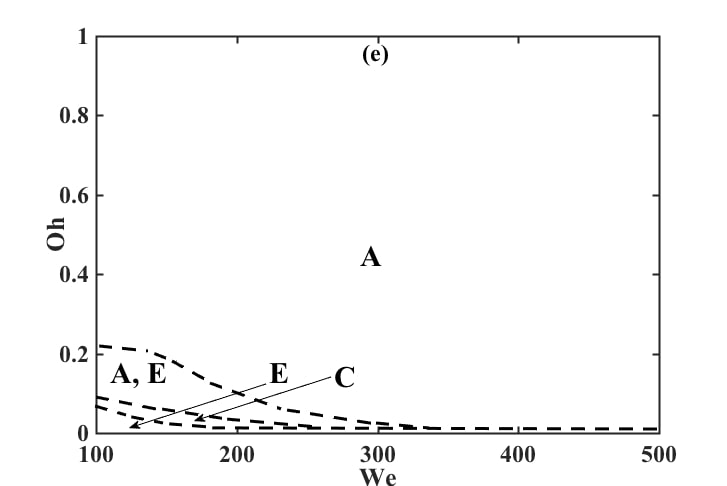}
\vskip -17pt
\caption*{} 
\end{subfigure}
\begin{subfigure}{0.245\textwidth}
 \includegraphics[width=1.1\linewidth, height=0.9\linewidth]{Fig5e.jpg}
 \vskip -17pt
 \caption*{} 
\end{subfigure}
\begin{subfigure}{0.245\textwidth}
\includegraphics[width=1.1\linewidth, height=0.9\linewidth]{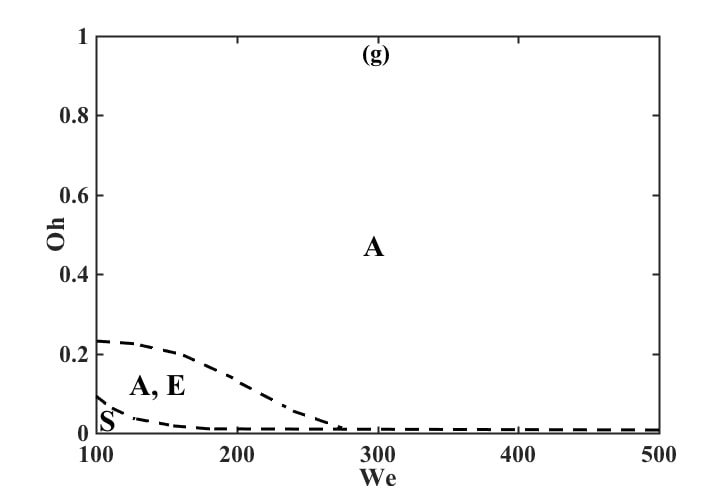}
\vskip -17pt
\caption*{} 
\end{subfigure}
\begin{subfigure}{0.245\textwidth}
 \includegraphics[width=1.1\linewidth, height=0.9\linewidth]{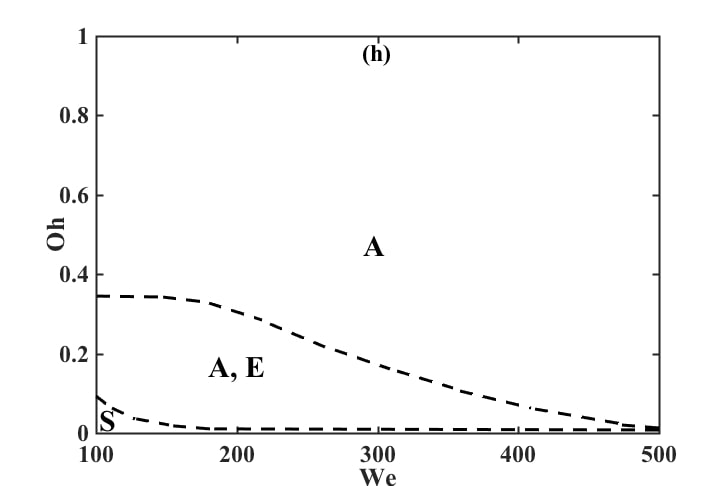}
 \vskip -17pt
 \caption*{} 
\end{subfigure}
\begin{subfigure}{0.245\textwidth}
\includegraphics[width=1.1\linewidth, height=0.9\linewidth]{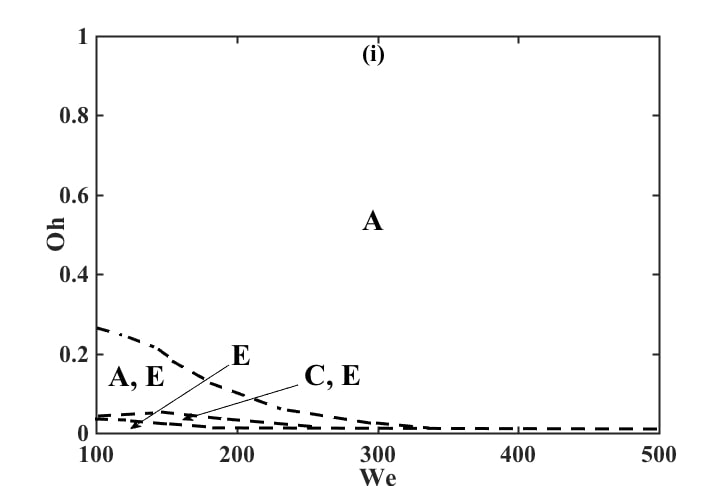}
\vskip -17pt
\caption*{} 
\end{subfigure}
\begin{subfigure}{0.245\textwidth}
 \includegraphics[width=1.1\linewidth, height=0.9\linewidth]{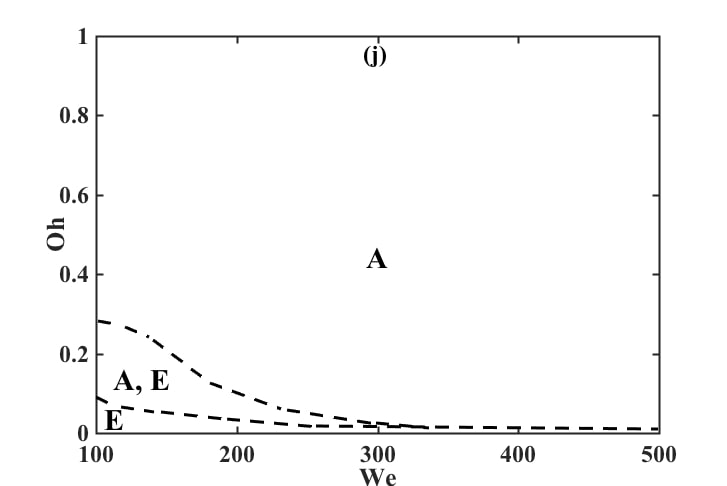}
 \vskip -17pt
 \caption*{} 
\end{subfigure}
\begin{subfigure}{0.245\textwidth}
 \includegraphics[width=1.1\linewidth, height=0.9\linewidth]{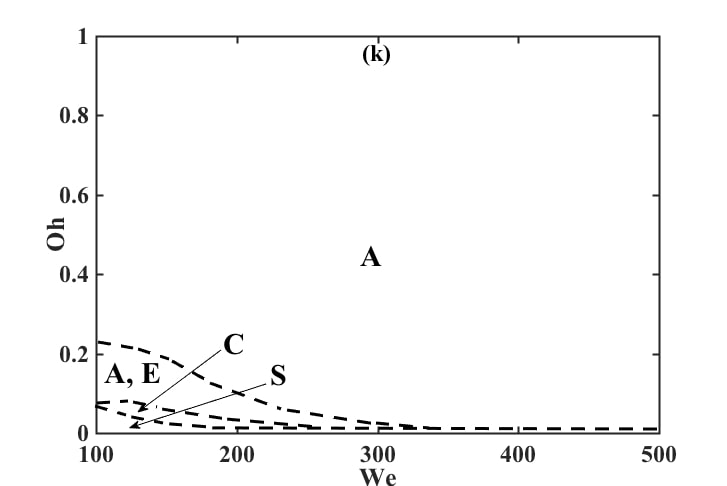}
\vskip -17pt
\caption*{} 
\end{subfigure}
\begin{subfigure}{0.245\textwidth}
  \includegraphics[width=1.1\linewidth, height=0.9\linewidth]{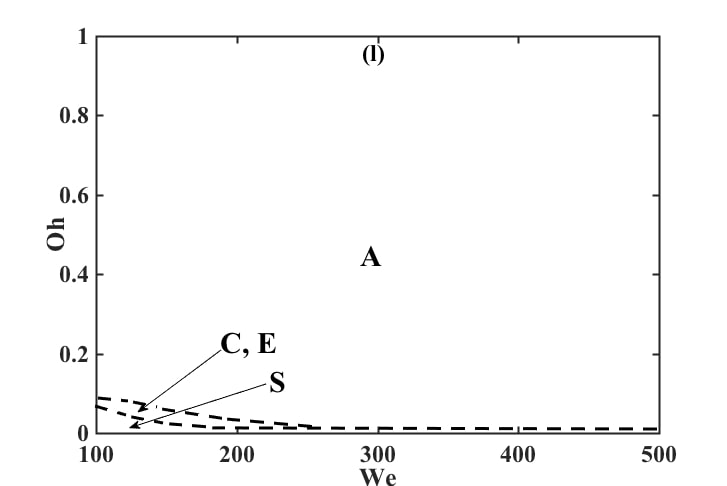}
 \vskip -17pt
 \caption*{} 
\end{subfigure}
\caption{Viscoelastic slender jet flow stability phase diagram in the $Oh-We$ parametric space at [(a)-(d)] $\nu=0.01$, $\epsilon=0.01$, [(e)-(h)] $\nu=0.99$, $\epsilon=0.01$, and [(i)-(l)] $\nu=0.01$, $\epsilon=0.99$; at spatiotemporal locations: $t=0.001, z=0.05$ (first column), $t=10.0, z=0.05$ (second column), $t=0.001, z=20.0$ (third column) and $t=10.0, z=20.0$ (fourth column). The regions {\bf S}, {\bf C}, and {\bf A} are denoted by temporally stable, convectively unstable and absolutely unstable regions, respectively. The domains outline by ({\bf A}, {\bf E}) and ({\bf C}, {\bf E}) are those where both the unstable and evanescent modes (denoted by {\bf E}) are found.}
\label{fig:Fig5}
\end{figure}

To explore the nature of these instabilities, we compute the boundaries of the temporally stable regions ({\bf S}), convective instabilities ({\bf C}), evanescent modes ({\bf E}) and absolute instabilities ({\bf A}) within a selected range of the flow-elasticity parameter space, i.~e., $Oh \in [0, 1.0], We \in [100, 500]$ and $\nu=0.01, \epsilon=0.01$ (first row in Fig.~\ref{fig:Fig5}), $\nu=0.99, \epsilon=0.01$ (second row in Fig.~\ref{fig:Fig5}) and $\nu=0.01, \epsilon=0.99$ (third row in Fig.~\ref{fig:Fig5}) and at spatiotemporal locations $(t, z)$: 0.001, 0.05 (first column), 10.0, 0.05 (second column), 0.001, 20.0 (third column) and 10.0, 20.0 (fourth column). Two observations are noteworthy in the phase diagrams. First, note that the impact of surface force stabilization at infinitesimally small values of $Oh$ is universally evident, i.~e., the presence of either convectively unstable or time-asymptotically stable region for $Oh < 10^{-3}$. The presence of either convectively unstable or evanescent modes for $Oh < 10^{-3}$, guarantees the absence of topological transitions leading to jet breakup~\cite{Duprat2007}. This observation is further corroborated by the in vitro studies by McKinley~\cite{Bhat2010} who concluded that significant inertia (alternatively, a small but critical value of $Oh$) is needed to induce satellite bead formation and breakup. Secondly, note the appearance of temporally stable region at larger values of $Oh$ (Fig.~\ref{fig:Fig5}c): a finding supported by previous in silico studies~\cite{Bhat2009} confirming the existence of a stable BOAS structure at significant inertia, for fluids exhibiting non-affine motion and modeled using the PTT network model. We conclude by noting that both the capillary and the inertial forces are instrumental in circumventing the topological transitions leading to breakup.

\section{Conclusions}\label{sec:Conclusions}
This investigation addresses the two-dimensional linear, spatiotemporal stability of viscoelastic slender jets moderately concentrated polymeric liquids, for low to moderate Ohnesorge number ($Oh$) and high values of Weissenberg number ($We$) to identify the regions of topological transition leading to pinch-off. The details of the slender body viscoelastic jet flow model, including the governing equations and the linear PTT model for the base constitutive relation for the extra elastic stress tensor; a quasi-stationary time-asymptotic solution of the model (or `mean flow'); the linearization around the mean flow and the numerical procedure for finding the absolute growth rate curves are outlined in \S \ref{subsec:GE}, \S \ref{subsec:MF}, \S \ref{subsec:LA} and \S \ref{subsec:NM}, respectively. The the spatiotemporal stability analyses (\S \ref{subsec:STI}) highlights the impact of finite stresses via strain-hardening and finite-time pinch-off, while the phase diagrams (\S \ref{subsubsec:FE}) divulge the influence of (a) capillary force stabilization at infinitesimally small values of $Oh$, and (b) inertial stabilization at significantly larger values of $Oh$.

Although this study provides an improved understanding of the effect of the finite extensibility of a real polymer chain on the breakup dynamics, a number of simplifying assumptions were made, including the absence of any entanglement effects for moderately concentrated polymeric liquids as well as no merging and draining of satellite drops at low viscosities or high surface tension ($Oh \ll 1 \ll We$)~\cite{Shamardi2018}. Another phenomena which is absent in our analysis, is the generation of secondary filaments via the `recoil' process~\cite{Panahi2020}. The absence of any coupled visco-elasto-capilliary effects as well as the development of singularity (or undefined slope) at points where the thread joins the drop~\cite{Keshavarz2016}, (and thus invalidating the current slender jet model) prompts us to explore the numerical solution of the full three-dimensional axisymmetric PTT equations capturing this complicated topological transition, in future.
\vskip 5pt
\noindent \textbf{Acknowledgments} T. C., D. B. and S.S. acknowledges the financial support of the Grant CSIR 09/1117(0012)/2020-EMR-I; CSIR 09/1117(0004)/2017-EMR-I and DST ECR/2017/000632, respectively.

\vskip 5pt
\noindent \textbf{Data availability} The data that support the findings of this study are available from the corresponding author upon reasonable request.

\vspace{-0.8cm}
\appendix
\section{Viscoelastic dispersion relation}  \label{sec:appendix}
The expression for the viscoelastic dispersion relation outlined in \S \ref{subsec:LA}, is given as,
\noindent 
\begin{align}
&D(\alpha, \omega) = \left[M^{11}_r + i M^{11}_i\right] \left[\left(-h_0 {\left(M^{33}_i\right)}^2 M^{33}_r - h_0 {\left(M^{33}_i\right)}^2 M^{44}_r - M^{22}_r {\left(M^{33}_i\right)}^2 + M^{22}_r M^{33}_r M^{44}_r -M^{22}_r 
\right.\right.\nonumber \\
&\left.\left. \epsilon^2 {A_{rr_0}} {A_{zz_0}}\right) - i \left(h_0 {\left(M^{33}_i\right)}^3 - h_0 M^{33}_i M^{33}_r M^{44}_r + h_0 M^{33}_i \epsilon^2 {A_{rr_0}} {A_{zz_0}} - M^{22}_r M^{33}_r M^{33}_i - 
M^{22}_r M^{44}_r M^{33}_i\right) \right] 
\nonumber \\
&+ \left[\alpha_r + i \alpha_i\right] \left[\dfrac{h_0^2}{2} \left[\left(-M^{12}_i {\left(M^{33}_i\right)}^2 + M^{12}_i M^{33}_r M^{44}_r - M^{12}_i \epsilon^2 {A_{rr_0}} {A_{zz_0}} + M^{12}_r M^{33}_i M^{33}_r + M^{12}_r M^{33}_i M^{44}_r\right)
\right. \right. \nonumber\\
& \left. \left.+  i \left(M^{12}_i M^{33}_i M^{33}_r + M^{12}_i M^{33}_i M^{44}_r + M^{12}_r {\left(M^{33}_i\right)}^2 - M^{12}_r M^{33}_r M^{44}_r + M^{12}_r \epsilon^2 {A_{rr_0}} {A_{zz_0}} \right) \right] - \left(\dfrac{1-\nu}{Re}\right) h_0^2 
\right. \nonumber \\
& \left. \left[\left( \left(-2 \alpha_r {A_{zz_0}} - \dfrac{2 \alpha_r}{\text{We}}\right) \left(-h_0 {\left(M^{33}_i\right)}^2 + M^{22}_r \epsilon {A_{rr_0}} + M^{22}_r M^{44}_r\right) + \left(2 \alpha_i {A_{zz_0}} + \dfrac{2 \alpha_i}{\text{We}}\right) \left(h_0 M^{33}_i  \epsilon {A_{rr_0}} + h_0 M^{33}_i
\right. \right. \right. \right.\nonumber \\
& \left. \left. \left. \left. M^{44}_r +M^{22}_r M^{33}_i\right) + \left(\alpha_r {A_{rr_0}} + \dfrac{\alpha_r}{\text{We}}\right) \left(h_0 {\left(M^{33}_i\right)}^2 - M^{22}_r M^{33}_r - M^{22}_r \epsilon {A_{zz_0}} \right) + \left(-\alpha_i {A_{rr_0}} -\dfrac{\alpha_i}{\text{We}}\right) \left( -h_0 M^{33}_i 
\right. \right. \right. \right.\nonumber \\
& \left. \left. \left. \left. 
M^{33}_r - h_0 M^{33}_i \epsilon {A_{zz_0}} - M^{22}_r M^{33}_i \right)\right) + i \left( \left(-2 \alpha_r {A_{zz_0}} - \dfrac{2 \alpha_r}{\text{We}}\right) \left(h_0 M^{33}_i \epsilon {A_{rr_0}} + h_0 M^{33}_i M^{44}_r +M^{22}_r M^{33}_i\right)+
\right. \right. \right.\nonumber \\
& \left. \left. \left. 
\left(2 \alpha_i {A_{zz_0}} + \dfrac{2 \alpha_i}{\text{We}}\right) \left(h_0 {\left(M^{33}_i\right)}^2 - M^{22}_r \epsilon {A_{rr_0}} - M^{22}_r M^{44}_r\right)   + \left(\alpha_r {A_{rr_0}} + \dfrac{\alpha_r}{\text{We}}\right) \left( -h_0 M^{33}_i M^{33}_r - h_0 M^{33}_i \epsilon {A_{zz_0}} 
\right. \right. \right. \right.\nonumber \\
& \left. \left. \left. \left.
- M^{22}_r M^{33}_i \right) + \left(-\alpha_i {A_{rr_0}} -\dfrac{\alpha_i}{\text{We}}\right) \left(- h_0 {\left(M^{33}_i\right)}^2 + M^{22}_r M^{33}_r + M^{22}_r \epsilon {A_{zz_0}} \right)
\right)\right]\right], \label{eqn:DRP}
\end{align}
\vskip 1pt
with its real part described by, \vskip 1pt
\noindent 
\begin{align}
&\text{Re}(D(\alpha, \omega)) = M^{11}_i \left[h_0 {\left(M^{33}_i\right)}^3 - h_0 M^{33}_i M^{33}_r M^{44}_r + h_0 M^{33}_i \epsilon^2 {A_{rr_0}} {A_{zz_0}} - M^{22}_r M^{33}_r M^{33}_i -  M^{22}_r M^{44}_r M^{33}_i\right] 
\nonumber \\
&+ M^{11}_r \left[-h_0 {\left(M^{33}_i\right)}^2 M^{33}_r - h_0 {\left(M^{33}_i\right)}^2 M^{44}_r - M^{22}_r {\left(M^{33}_i\right)}^2 + M^{22}_r M^{33}_r M^{44}_r - M^{22}_r \epsilon^2 {A_{rr_0}} {A_{zz_0}}\right] + \alpha_r \dfrac{h_0^2}{2} 
\nonumber \\
& \left[-M^{12}_i {\left(M^{33}_i\right)}^2 + M^{12}_i M^{33}_r M^{44}_r - M^{12}_i \epsilon^2 {A_{rr_0}} {A_{zz_0}} + M^{12}_r M^{33}_i M^{33}_r + M^{12}_r M^{33}_i M^{44}_r\right] - \alpha_i \dfrac{h_0^2}{2} \left[M^{12}_i M^{33}_i M^{33}_r
\right. \nonumber \\
& \left. + M^{12}_i M^{33}_i M^{44}_r + M^{12}_r {\left(M^{33}_i\right)}^2 - M^{12}_r M^{33}_r M^{44}_r + M^{12}_r \epsilon^2 {A_{rr_0}} {A_{zz_0}} \right] - \alpha_r \left(\dfrac{1-\nu}{Re}\right) h_0^2 \left[ \left(-2 \alpha_r {A_{zz_0}} - \dfrac{2 \alpha_r}{\text{We}}\right)
\right. \nonumber \\
& \left.
\left(-h_0 {\left(M^{33}_i\right)}^2 + M^{22}_r \epsilon {A_{rr_0}} + M^{22}_r M^{44}_r\right) + \left(2 \alpha_i {A_{zz_0}} + \dfrac{2 \alpha_i}{\text{We}}\right) \left(h_0 M^{33}_i \epsilon {A_{rr_0}} + h_0 M^{33}_i M^{44}_r +M^{22}_r M^{33}_i\right) + \left(\alpha_r
\right. \right. \nonumber \\
& \left. \left.
{A_{rr_0}} + \dfrac{\alpha_r}{\text{We}}\right) \left(h_0 {\left(M^{33}_i\right)}^2 - M^{22}_r M^{33}_r - M^{22}_r \epsilon {A_{zz_0}} \right) + \left(-\alpha_i {A_{rr_0}} -\dfrac{\alpha_i}{\text{We}}\right) \left( -h_0 M^{33}_i M^{33}_r - h_0 M^{33}_i \epsilon {A_{zz_0}} - M^{22}_r
\right. \right. \nonumber \\
& \left. \left.
M^{33}_i \right)\right] + \alpha_i \left(\dfrac{1-\nu}{Re}\right) h_0^2 \left[ \left(-2 \alpha_r {A_{zz_0}} - \dfrac{2 \alpha_r}{\text{We}}\right) \left(h_0 M^{33}_i \epsilon {A_{rr_0}} + h_0 M^{33}_i M^{44}_r +M^{22}_r M^{33}_i\right)+ \left(2 \alpha_i {A_{zz_0}} + \dfrac{2 \alpha_i}{\text{We}}\right)
\right. \nonumber \\
& \left.
\left(h_0 {\left(M^{33}_i\right)}^2 - M^{22}_r \epsilon {A_{rr_0}} - M^{22}_r M^{44}_r\right)   + \left(\alpha_r {A_{rr_0}} + \dfrac{\alpha_r}{\text{We}}\right) \left( -h_0 M^{33}_i M^{33}_r - h_0 M^{33}_i \epsilon {A_{zz_0}} - M^{22}_r M^{33}_i \right) + \left(-\alpha_i
\right. \right. \nonumber \\
& \left. \left.
{A_{rr_0}} -\dfrac{\alpha_i}{\text{We}}\right) \left(- h_0 {\left(M^{33}_i\right)}^2 + M^{22}_r M^{33}_r 
+ M^{22}_r \epsilon {A_{zz_0}} \right) \right], \label{eqn:DRP_real}
\end{align}
\vskip 1pt
and its imaginary part delineated as, \vskip 1pt
\noindent
\begin{align}
&\text{Im}(D(\alpha, \omega)) = M^{11}_r \left[-h_0 {\left(M^{33}_i\right)}^3 + h_0 M^{33}_i M^{33}_r M^{44}_r - h_0 M^{33}_i \epsilon^2 {A_{rr_0}} {A_{zz_0}} + M^{22}_r M^{33}_r M^{33}_i + M^{22}_r M^{44}_r M^{33}_i\right] 
\nonumber \\
& + M^{11}_i \left[- h_0 {\left(M^{33}_i\right)}^2 M^{33}_r - h_0 {\left(M^{33}_i\right)}^2 M^{44}_r -  M^{22}_r {\left(M^{33}_i\right)}^2 + M^{22}_r M^{33}_r M^{44}_r - M^{22}_r \epsilon^2 {A_{rr_0}} {A_{zz_0}}\right] - \alpha_i \dfrac{h_0^2}{2} \left[M^{12}_i 
\right. \nonumber \\
& \left. {\left(M^{33}_i\right)}^2 - M^{12}_i M^{33}_r M^{44}_r + M^{12}_i \epsilon^2 {A_{rr_0}} {A_{zz_0}} - M^{12}_r M^{33}_i M^{33}_r - M^{12}_r M^{33}_i M^{44}_r\right] + \alpha_r \dfrac{h_0^2}{2}\left[M^{12}_i M^{33}_i M^{33}_r + M^{12}_i
\right. \nonumber \\
& \left.
M^{33}_i M^{44}_r + M^{12}_r {\left(M^{33}_i\right)}^2 - M^{12}_r M^{33}_r M^{44}_r + M^{12}_r \epsilon^2 {A_{rr_0}} {A_{zz_0}} \right] + \alpha_i \left(\dfrac{1-\nu}{Re}\right) h_0^2 \left[ \left(-2 \alpha_r {A_{zz_0}} - \dfrac{2 \alpha_r}{\text{We}}\right) \left(h_0 {\left(M^{33}_i\right)}^2
\right. \right. \nonumber \\
& \left. \left.
- M^{22}_r \epsilon {A_{rr_0}} - M^{22}_r M^{44}_r\right) + \left(2 \alpha_i {A_{zz_0}} + \dfrac{2 \alpha_i}{\text{We}}\right) \left(- h_0 M^{33}_i \epsilon {A_{rr_0}} - h_0 M^{33}_i M^{44}_r - M^{22}_r M^{33}_i\right) + \left(\alpha_r {A_{rr_0}} + \dfrac{\alpha_r}{\text{We}}\right)
\right. \nonumber \\
& \left.
\left(-h_0 {\left(M^{33}_i\right)}^2 + M^{22}_r M^{33}_r + M^{22}_r \epsilon {A_{zz_0}} \right) + \left(-\alpha_i {A_{rr_0}} -\dfrac{\alpha_i}{\text{We}}\right) \left( h_0 M^{33}_i M^{33}_r + h_0 M^{33}_i \epsilon {A_{zz_0}} + M^{22}_r M^{33}_i \right)\right] - \alpha_r
\nonumber \\
& 
\left(\dfrac{1-\nu}{Re}\right) h_0^2 \left[ \left(-2 \alpha_r {A_{zz_0}} - \dfrac{2 \alpha_r}{\text{We}}\right) \left(h_0 M^{33}_i \epsilon {A_{rr_0}} + h_0 M^{33}_i M^{44}_r +M^{22}_r M^{33}_i\right)+ \left(2 \alpha_i {A_{zz_0}} + \dfrac{2 \alpha_i}{\text{We}}\right)\left(h_0 {\left(M^{33}_i\right)}^2 
\right. \right.\nonumber \\
& \left. \left.
- M^{22}_r \epsilon {A_{rr_0}} - M^{22}_r M^{44}_r\right)   + \left(\alpha_r {A_{rr_0}} + \dfrac{\alpha_r}{\text{We}}\right) \left( -h_0 M^{33}_i M^{33}_r - h_0 M^{33}_i \epsilon {A_{zz_0}} - M^{22}_r M^{33}_i \right) + \left(-\alpha_i {A_{rr_0}} -\dfrac{\alpha_i}{\text{We}}\right) 
\right. \nonumber \\
& \left.
\left(- h_0 {\left(M^{33}_i\right)}^2 + M^{22}_r M^{33}_r + M^{22}_r \epsilon {A_{zz_0}} \right)
\right]. \label{eqn:DRP_imag}
\end{align}
%
%
%
The expressions utilized in equations \eqref{eqn:DRP}-\eqref{eqn:DRP_imag} are,
\begin{align}
&M^{11} = \left(\omega_i {h_0}^2 - \alpha_i u_0 {h_0}^2 + \dfrac{d u_0}{dz} {h_0}^2 + \left(\dfrac{3 \nu}{Re}\right) \left( {\alpha_r}^2 - {\alpha_i}^{2}\right) {h_0}^2\right) + i \left(-\omega_r {h_0}^2 + \alpha_r u_0 {h_0}^2 + \left(\dfrac{6 \nu}{Re}\right) \alpha_r \alpha_i {h_0}^2\right) \nonumber \\
& M^{12} = \left(2 u_0 \dfrac{d u_0}{dz} h_0 + \dfrac{\alpha_i}{Oh^2 Re} + \alpha_i \left(\dfrac{6 \nu}{Re}\right) \dfrac{d u_0}{dz} h_0 + 2 \alpha_i \dfrac{\left(1-\nu\right)}{Re} h_0 \left({A_{zz_0}}-{A_{rr_0}}\right) + \left(\dfrac{(\alpha_i^3 - 3 \alpha_r^2 \alpha_i) h_0^2}{Oh^2 Re}\right)\right) \nonumber \\
&\quad \quad+ i \left( -\dfrac{\alpha_r}{Oh^2 Re} - \alpha_r \left(\dfrac{6 \nu}{Re}\right) \dfrac{d u_0}{dz} h_0 - 2 \alpha_r \dfrac{\left(1-\nu\right)}{Re} h_0 \left({A_{zz_0}}-{A_{rr_0}}\right) + \left(\dfrac{(\alpha_r^3 - 3 \alpha_r \alpha_i^2) h_0^2}{Oh^2 Re}\right)\right) \nonumber \\
&M^{13} = -M^{14} = \left(\alpha_i \dfrac{\left(1-\nu\right)}{Re} {h_0}^2\right) + i \left(-\alpha_r \dfrac{\left(1-\nu\right)}{Re} {h_0}^2\right) \nonumber\\
&M^{21} = \left( \dfrac{- \alpha_i h_0^2}{2}\right) + i  \left(\dfrac{\alpha_r h_0^2}{2}\right) \nonumber \\
& M^{22} = \left(\dfrac{d h_0}{dt}  + \omega_i h_0 + \dfrac{d u_0}{dz} h_0 - \alpha_i h_0 u_0\right) + i \left( -\omega_r h_0 + \alpha_r h_0 u_0\right) \nonumber \\
& M^{31} = \left(2 \alpha_i {A_{zz_0}} + 2 \left(\dfrac{\alpha_i}{\text{We}}\right)\right) + i \left( -2 \alpha_r {A_{zz_0}} - 2 \left(\dfrac{\alpha_r}{\text{We}}\right)\right) \nonumber \\
& M^{33} = \left(\omega_i - \alpha_i u_0 - 2 \dfrac{d u_0}{dz} + \dfrac{1}{\text{We}} + 2 \epsilon {A_{zz_0}} + \epsilon {A_{rr_0}}\right) + i \left(-\omega_r + \alpha_r u_0\right) \nonumber \\
& M^{34} = \epsilon {A_{zz_0}} \nonumber \\
& M^{41} = \left(- \alpha_i {A_{rr_0}} -\left(\dfrac{\alpha_i}{\text{We}}\right)\right) + i \left(\alpha_r {A_{rr_0}} + \left(\dfrac{\alpha_r}{\text{We}}\right)\right) \nonumber \\
& M^{43} = \epsilon {A_{rr_0}} \nonumber \\
& M^{44} = \left(\omega_i - \alpha_i u_0 + \dfrac{d u_0}{dz} + \dfrac{1}{\text{We}} + \epsilon {A_{zz_0}} + 2 \epsilon {A_{rr_0}}\right) + i \left(-\omega_r + \alpha_r u_0\right)
\end{align}

\bibliographystyle{unsrt}

\end{document}